\documentclass[useAMS,usenatbib]{mn2e}
\usepackage{amssymb,amsmath,graphicx,color}

\newcommand{\gn}{\ensuremath{\Gamma_\mathrm{0}}}
\newcommand{\tj}{\ensuremath{\theta_\mathrm{j}}}
\newcommand{\tv}{\ensuremath{\theta_\mathrm{v}}}
\newcommand{\ts}{\ensuremath{\theta_\mathrm{s}}}
\newcommand{\tl}{\ensuremath{\theta_\mathrm{L}}}
\newcommand{\pl}{\ensuremath{\phi_\mathrm{L}}}
\newcommand{\rph}{\ensuremath{R_\mathrm{ph}}}
\newcommand{\I}{\ensuremath{\mathcal{I}}}
\newcommand{\Q}{\ensuremath{\mathcal{Q}}}
\newcommand{\U}{\ensuremath{\mathcal{U}}}
\newcommand{\V}{\ensuremath{\mathcal{V}}}
\newcommand{\tn}{\ensuremath{\theta_\mathrm{0}}}
\newcommand{\pn}{\ensuremath{\phi_\mathrm{0}}}
\newcommand{\tsc}{\ensuremath{\theta_\mathrm{sc}}}
\newcommand{\psc}{\ensuremath{\phi_\mathrm{sc}}}

\newcommand{\coord}{\ensuremath{\mathrm{C}}}
\newcommand{\coordsup}[1]{\ensuremath{\mathrm{C^{#1}}}}
\newcommand{\coordsub}[1]{\ensuremath{\mathrm{C_{#1}}}}
\newcommand{\coordsubsup}[2]{\ensuremath{\mathrm{C_{#1}^{#2}}}}
\newcommand{\vect}[1]{\ensuremath{\mathbf{#1}}}
\newcommand{\vectsub}[2]{\ensuremath{\mathbf{#1}_\mathrm{#2}}}
\newcommand{\vectsup}[2]{\ensuremath{\mathbf{#1}^\mathrm{#2}}}
\newcommand{\vectsubsup}[3]{\ensuremath{\mathbf{#1}_{#2}^\mathrm{#3}}}
\newcommand{\vecthat}[1]{\ensuremath{\mathbf{\hat{#1}}}}
\newcommand{\vectsubhat}[2]{\ensuremath{\mathbf{\hat{#1}}_\mathrm{#2}}}
\newcommand{\vectsuphat}[2]{\ensuremath{\mathbf{\hat{#1}}^\mathrm{#2}}}
\newcommand{\fvect}[1]{\ensuremath{\mathrm{#1}}}
\newcommand{\fvectsub}[2]{\ensuremath{\mathrm{{#1}_{#2}}}}
\newcommand{\fvectsup}[2]{\ensuremath{\mathrm{{#1}^{#2}}}}
\newcommand{\fvectsubsup}[3]{\ensuremath{\mathrm{{#1}_{#2}^{#3}}}}

\newcommand{\D}[2]{\ensuremath{\frac{\mathrm{d}#1}{\mathrm{d}#2}}}

\title[Polarization of photospheric emission]{Polarization properties of photospheric emission from relativistic, collimated outflows}
\author[C. Lundman, A. Pe'er and F. Ryde]{C. Lundman$^{1,2}$\thanks{E-mail:
clundman@particle.kth.se (CL); a.peer@ucc.ie (AP); fryde@kth.se (FR)}, A. Pe'er$^{3}$\footnotemark[1] and F. Ryde$^{1,2}$\footnotemark[1] \\
$^{1}$Department of Physics, KTH Royal Institute of Technology, AlbaNova, SE-106 91 Stockholm, Sweden \\
$^{2}$The Oskar Klein Centre for Cosmoparticle Physics, AlbaNova, SE-106 91 Stockholm, Sweden \\
$^{3}$Physics Department, University College Cork, Cork, Ireland}
\begin{document}

\date{Accepted --}

\pagerange{\pageref{firstpage}--\pageref{lastpage}} \pubyear{2013}

\maketitle

\label{firstpage}

\begin{abstract}
We consider the polarization properties of photospheric emission originating in jets consisting of a highly relativistic core of opening angle $\tj$ and Lorentz factor $\gn$, and a surrounding shear layer where the Lorentz factor is decreasing as a power law of index $p$ with angle from the jet axis. We find significant degrees of linear polarization for observers located at viewing angles $\tv \gtrsim \tj$. In particular, the polarization degree of emission from narrow jets ($\tj \approx 1/\gn$) with steep Lorentz factor gradients ($p \gtrsim 4$) reaches $\sim 40 \%$. The angle of polarization may shift by $\pi/2$ for time-variable jets. The spectrum below the thermal peak of the polarized emission appears non-thermal due to aberration of light, without the need for additional radiative processes or energy dissipation. Furthermore, above the thermal peak a power law of photons forms due to Comptonization of photons that repeatedly scatter between regions of different Lorentz factor before escaping. We show that polarization degrees of a few tens of percent and broken power law spectra are natural in the context of photospheric emission from structured jets. Applying the model to gamma-ray bursts, we discuss expected correlations between the spectral shape and the polarization degree of the prompt emission.
\end{abstract}


\section{Introduction}
\label{sect:introduction}

Relativistic jets are found in a variety of astrophysical objects. In spite of large uncertainties, it is clear that in several types of objects the jets are optically thick at the launching region. This applies to the innermost regions of the jets in X-ray binaries and active galactic nuclei. However, the foremost examples are the ultra-relativistic jets responsible for producing gamma-ray bursts (GRBs), which are optically thick up to distances several order of magnitude larger than the jet base. As the jets expand and transition to transparency, they release internally trapped photons as photospheric emission.

Although no consensus has been reached, evidence are accumulating that photospheric emission may play a significant role, and in a few cases dominate the prompt GRB emission. Optically thin synchrotron emission originating from internal shocks \citep{ReeMes:1994} has been the most common interpretation for many years, due to the common broken power law shape of the observed spectra. However, the model faces severe challenges. First, basic synchrotron theory can not explain the steep spectrum observed below the peak energy in a substantial fraction of GRBs \citep{PreEtAl:1998, KanEtAl:2006, NavEtAl:2011, GolEtAl:2012}. Second, it provides no natural explanation for the clustering of observed peak energies at a few hundred keV. Third, the energy budget for the prompt emission consists of the relative kinetic energy dissipated in the internal shocks, which leads to efficiency problems. These have led to renewed interest in alternative prompt emission models.

Evidence for a photospheric origin of at least part of the prompt emission have now been found in numerous long GRBs (see e.g. \citealt{Ryd:2004, Ryd:2005, RydPee:2009, RydEtAl:2010, GuiEtAl:2011, AxeEtAl:2012, IyyEtAl:2013, GhiPesGhi:2013}), as well as in the short GRB 120323A \citep{GuiEtAl:2013}. In the case of GRB 090902B, almost Planck-like photospheric emission appears to dominate the observed emission \citep{RydEtAl:2010}, and the narrowly peaked spectrum is observed to broaden over the duration of the burst into a more typical, smoothly broken power law shape \citep{RydEtAl:2011}. Indeed, in recent years it was realized from a theoretical perspective that while photospheric emission can be Planckian, the spectrum emitted at the photosphere is in general expected to be broadened \citep{ReeMes:2005, PeeMesRee:2005, PeeMesRee:2006, Gia:2006, Pee:2008, Bel:2010, PeeRyd:2011, VurBelPou:2011, LunPeeRyd:2013, ItoEtAl:2013}. A photospheric origin of the prompt emission naturally explains the clustering of observed peak energies, as the observed temperature is insensitive to the outflow parameters. Steep spectral shapes below the peak energy are also easily accomodated within the photospheric model. On the other hand, the presence of high energy emission ($\gtrsim 100 \, \mathrm{MeV}$) is hard to explain with a simple photospheric model due to photon-photon pair production. Therefore, the origin of the prompt emission is still debatable.

Polarization measurements of the prompt emission offer additional information that may help resolve the issue. The first claim of detection of polarized prompt emission was made by \citet{CobBog:2003} using {\it RHESSI} data of GRB 021206. The reported linear polarization degree was high ($\Pi = 80 \% \pm 20 \%$), however the claims could not be verified by independent analysis of the data \citep{RutFox:2004, WigEtAl:2004}. Using {\it INTEGRAL} data in the energy range 100 keV $-$ 1 MeV, high polarization degrees has been measured in GRB 041219A (\citealt{KalEtAl:2007}, \citealt{McGEtAl:2007} and \citealt{GotEtAl:2009} reports $\Pi = 98 \% \pm 33 \%$, $\Pi = 63^{+31 \%}_{-30 \%}$ and $\Pi = 43 \% \pm 25 \%$ respectively) and GRB 061121 (\citealt{McGEtAl:2009} and \citealt{GotEtAl:2013} reports $\Pi = 29^{+25 \%}_{-26 \%}$ and $\Pi > 30 \%$ respectively), although instrumental systematics could not be ruled out. More recently, \citet{YonEtAl:2011, YonEtAl:2012} measured polarization degrees of $\Pi = 25 \% \pm 15 \%$ and $\Pi = 31 \% \pm 21 \%$ in two different time intervals of GRB 100826A, $\Pi = 70 \% \pm 22 \%$ in GRB 110301A and $\Pi = 84^{+16 \%}_{-28 \%}$ in GRB 110721A using data from the {\it GAP} instrument. Unfortunately, all polarization measurements of prompt emission to date suffer from low photon statistics, leading to large uncertainties in the measurements.

The high polarization degree claimed by \citet{CobBog:2003} encouraged a large theoretical effort on the polarization predictions of various competing prompt emission models. These include synchrotron emission in a globally ordered magnetic field \citep{Gra:2003, GraKon:2003, NakPirWax:2003, LyuParBla:2003}, synchrotron emission in a random magnetic field within the plane orthogonal to the local expansion direction \citep{Wax:2003, GraKon:2003}, Compton drag (upscattering of a background photon field by the jet, \citealt{LazEtAl:2004}) and Compton sailing (reflection of photons on the surrounding gas that collimates the jet, \citealt{EicLev:2003, LevEic:2004}). For reviews of the models above, see \citet{Laz:2006}, \citet{TomEtAl:2009} and \citet{Tom:2013}.

The previously mentioned models assume the prompt emission to originate in a transparent region of the jet. As the number of polarization measurements is growing, and the observational evidence for photospheric emission is accumulating, quantitative predictions regarding the polarization properties of photospheric emission is needed for comparison to data.

When observing emission from the whole emitting region simultaneously, the observed polarization signal can vanish, even though the local fluid elements emit polarized emission. This happens for all spatially unresolved sources with rotational symmetry around the line-of-sight (LOS). Jets, by their very nature, have a lateral structure (i.e. angle dependent outflow properties). Therefore, a natural way to break the symmetry is by observing the jet off-axis. Hydrodynamical simulations of GRB jets propagating through the stellar envelope (e.g. \citealt{ZhaWooMac:2003, MorLazBeg:2007, MizNagAoi:2011}) show the development of a central jet core, with outflow properties that are approximately constant with angle from the jet axis, and a shear layer, where the core connects to the surrounding stellar gas. The presence of the shear layer is omitted in many theoretical works that consider radiative processes, and the jets are commonly assumed to be either top-hats or spherically symmetric outflows.

The properties of the spectrum and polarization of photospheric emission from a structured jet is different from those of a spherical outflow. The deviation is particularly significant, but not limited to, when the shear layer is within the field of view. In particular, emission from the shear layer leads to a softer spectrum below the observed peak energy, as compared to photospheric emission from a spherical wind \citep{LunPeeRyd:2013}. In fact, depending on the typical angular widths of the jet core and shear layer, observing photospheric emission only from the jet core may be rare. This could explain the scarcity of GRBs with spectra similar to GRB 090902B (whose spectrum appeared as a quasi-thermal component, see \citealt{RydEtAl:2010, RydEtAl:2011}), which may have been observed on axis. Another consequence is that the photosphere would be asymmetric in most GRBs since most observers are located significantly off-axis, and therefore an inherent potential for producing polarized emission from the photosphere exists.

In order to isolate the effects of jet geometry and observer viewing angle on the resulting photospheric emission, we consider steady, non-dissipative jets, which cool passively through adiabatic expansion. Therefore, no contributions to the emission from other emission processes such as synchrotron emission are considered. We emphasize that while heating of the jet can modify the observed spectrum, it has little impact on the transfer of photon number in the jet \citep{Bel:2011}. Therefore, the polarization properties of the emission presented in this work are expected to be valid also for heated jets as long as the dissipation does not significantly affect the jet dynamics.

We assume that the jet develops an angle dependent baryon loading through the interaction with its surroundings. The baryon loading per solid angle is considered constant within the jet core, and then increase as a power law with angle in the shear layer. The baryon loading naturally leads to an angle dependent bulk Lorentz factor in the jet coasting regime. The observed polarization properties of the emission released at the photosphere is then computed for jets with different core and shear layer widths.

We show below that emission from the photosphere can be significantly polarized for observers located at viewing angles $\tv \gtrsim \tj$, where $\tj$ is the jet core opening angle. The polarization degree is largest for narrow jets ($\tj \gn \approx 1$, where $\gn$ is the jet core Lorentz factor) with a large Lorentz factor gradient in the shear layer, since this combination maximizes the asymmetry of the photospheric emitting region. As we show here, for such jets the polarization degree reaches up to $\sim 40 \%$ at viewing angles where the jet is still clearly observable. All jets within the considered parameter space produce at least a few percent of polarized emission at large viewing angles. For $\tv \gtrsim \tj$, geometrical broadening of the spectum below the peak energy occurs as the emission originating from the jet core and shear layer experience different Doppler boosts (see \citealt{LunPeeRyd:2013} for details on geometrical broadening). Our model therefore predicts a correlation between the low energy photon index and the polarization degree of the prompt emission. Moreover, as shown by \citet{LunPeeRyd:2013} and \citet{ItoEtAl:2013}, photons that scatter repeatedly within the shear layer may obtain energies above the local temperature. This process is efficient in jets with steep Lorentz factor gradients. Jets with narrow shear layers therefore give rise to both highly polarized emission and efficient Comptonization. Assuming the powerlaw tail commonly observed above the peak energy in GRBs originates from Comptonized shear layer photons, a correlation is expected between the degree of polarization and the strength of the emission above the thermal peak. Although we consider outflow parameters characterizing GRB jets in this work, the results obtained here are general and can be applied to other optically thick, relativistic outflows.

This paper is organized as follows. In \S \ref{sect:model} the model is presented in detail. A simplified analytical calculation of the expected polarization properties of photospheric emission from narrow jets is presented in \S \ref{sect:analytical}. The radiative transfer in non-dissipative fireball jets is analyzed using a Monte Carlo code, which is briefly explained in \S \ref{sect:MC}. The polarization results obtained from the simulation are presented in \S \ref{sect:results}. In \S \ref{sect:discussion} we discuss the results and compare them to polarization predictions of synchrotron emission. The details on the numerical integration of the simplified model of \S \ref{sect:analytical} is presented in Appendix \ref{sect:appendix analytic solution}. The details of simulating Lorentz transformations and scatterings of polarized emission are given in Appendix \ref{sect:appendix simulation}, while the consequences of the emitting region asymmetry are discussed in Appendix \ref{sect:appendix emitting region}.


\section{Polarized emission from a structured jet}
\label{sect:model}

As will be explained in detail below (\S \ref{subsect:compton}), there are two basic requirements for producing polarized emission through Compton scattering in a spatially unresolved outflow. First, the comoving intensity streaming through a local, emitting fluid element must be anisotropic. In the context of photospheric emission from astronomical jets this requires the outflow to be expanding, and not dominated by radiation at the photosphere \citep{Bel:2011}. Second, the jet has to have some lateral structure while the observer is located off-axis, so that the observed emitting region is not symmetric around the LOS. The most general model would therefore only make assumptions on these properties of the jet close to the photosphere. In order to produce quantitative results we here make the additional assumption of GRB ``fireball'' dynamics and perform radiative transfer simulations of the fireball emission. We stress that the polarization properties of the emission are not sensitive to the specific fireball parameters such as the value of the isotropic equivalent luminosity of the jet, or the size of the base of the outflow.

\subsection{The structured fireball model}

We consider the jet interaction with the surrounding gas to be summarized as an angle dependent baryon loading, $\mathrm{d}\dot{M}/\mathrm{d}\Omega = \mathrm{d}\dot{M}(\theta)/\mathrm{d}\Omega$, where $\mathrm{d}\dot{M}(\theta)/\mathrm{d}\Omega$ is the mass outflow rate per solid angle, and $\theta$ is the angle to the jet axis of symmetry. For example, in the collapsar model the surrounding gas is the stellar envelope, which confines the jet. The baryon poor jet core accelerates to a larger radius before saturating than the jet edge, which carries more baryons. This process leads to the development of an angle dependent outflow Lorentz factor of the jet plasma.

Fluid elements in relativistic, radially expanding outflows that are separated by an angle $1/\Gamma$, where $\Gamma$ is the Lorentz factor of the outflow, are out of causal contact. When the outflow Lorentz factor grows large, $1/\Gamma$ is much smaller than unity. We therefore make the simplifying assumption that each local fluid element propagates radially, and that the dynamics of a given fluid element follows that of a fluid element in a non-dissipative spherical fireball with the same fluid properties.

We consider the outflow close to the photosphere to be in the coasting phase, where the Lorentz factor has saturated to a value equal to the dimensionless entropy of the outflow,

\begin{equation}
\eta(\theta) = \frac{\mathrm{d}L(\theta)/\mathrm{d}\Omega}{c^2 \mathrm{d}\dot{M}(\theta)/\mathrm{d}\Omega},
\label{eq:eta}
\end{equation}

\noindent where $\mathrm{d}L(\theta)/\mathrm{d}\Omega$ is the luminosity per solid angle of the jet and $c$ is the speed of light.

The optical depth between two points in the jet separated by a distance $\mathrm{d}s$ equals

\begin{equation}
\mathrm{d}\tau = \Gamma(\theta)[1 - \beta(\theta) \cos\theta_\mathrm{rel}] n^\prime(r, \, \theta) \sigma \mathrm{d}s
\label{eq:tau}
\end{equation}

\noindent where $\theta_\mathrm{rel}$ is the angle between the photon propagation direction and the local outflow propagation direction, $\beta = v/c = \sqrt{1-\Gamma^{-2}}$ is the outflow speed in units of the speed of light and $\sigma$ is the scattering cross section. The comoving electron number density is

\begin{equation}
n^\prime(r, \, \theta) = \frac{1}{r^2 m_\mathrm{p}c^2 \beta (\theta) \Gamma(\theta)} \D{\dot{M}(\theta)}{\Omega},
\end{equation}

\noindent where $r$ is the distance from the center of the outflow, $m_\mathrm{p}$ is the proton mass and the assumption of radial motion has been used.

As the jet expands, the mean free path of the trapped fireball photons increases. The observed photons escape the outflow in a volume surrounding the photosphere, which is defined as the surface from which the optical depth for a photon that propagates towards the observer equals unity. Since the outflow is moving with a speed comparable to $c$, the optical depth is strongly dependent on the angle between the photon propagation direction and the local velocity field (equation \ref{eq:tau}, see \citealt{AbrNovPac:1991, Pee:2008, Bel:2011, LunPeeRyd:2013} for detailed discussions of this effect).

For simplicity, we consider the luminosity per solid angle of the central engine to be angle independent within the jet core and shear layer, $\mathrm{d}L/\mathrm{d}\Omega = L/4\pi$, where $L$ is the total, isotropic equivalent luminosity of the central engine\footnote{This statement does {\it not} imply that the luminosity of radiation emitted by the jet is constant with respect to observer viewing angle. The photospheric radius in the shear layer is larger than in the jet core \citep{Pee:2008, LunPeeRyd:2013}. Therefore, emission released by the shear layer has lost more energy to adiabatic expansion than emission released by the jet core, and an increase of the viewing angle leads to a decrease of the observed luminosity.}. Therefore, the angle dependence of the saturated Lorentz factor is uniquely determined by the angle dependence of the baryon loading.

The comoving temperature of the jet plasma is determined by the size of the jet base region, $r_\mathrm{0}$, and the central engine luminosity. The temperature at the base is $T_\mathrm{0} = (L/4 \pi r_\mathrm{0}^2 a c)^{1/4}$, where $a$ is the radiation constant. The saturation radius, above which the outflow is coasting, equals $r_\mathrm{s}(\theta) = \eta(\theta) r_\mathrm{0}$. The comoving temperature of the outflow at angle $\theta$ and radius $r > r_\mathrm{s}(\theta)$ is then

\begin{equation}
T^\prime(r, \theta) = T_\mathrm{0} \frac{r_\mathrm{0}}{r_\mathrm{s} (\theta)} \left(\frac{r_\mathrm{s} (\theta)}{r}\right)^{2/3}.
\label{eq:comoving temp}
\end{equation}

The photon emission rate from the central engine is obtained by noting that photons dominate the energy density at the jet base, and the average photon energy is $2.7 k T_\mathrm{0}$ where $k$ is the Boltzmann constant. Therefore, $\mathrm{d}\dot{N}_\gamma/\mathrm{d}\Omega = L/(4\pi \cdot 2.7 k T_\mathrm{0})$.

\begin{figure}
\includegraphics[width=\linewidth]{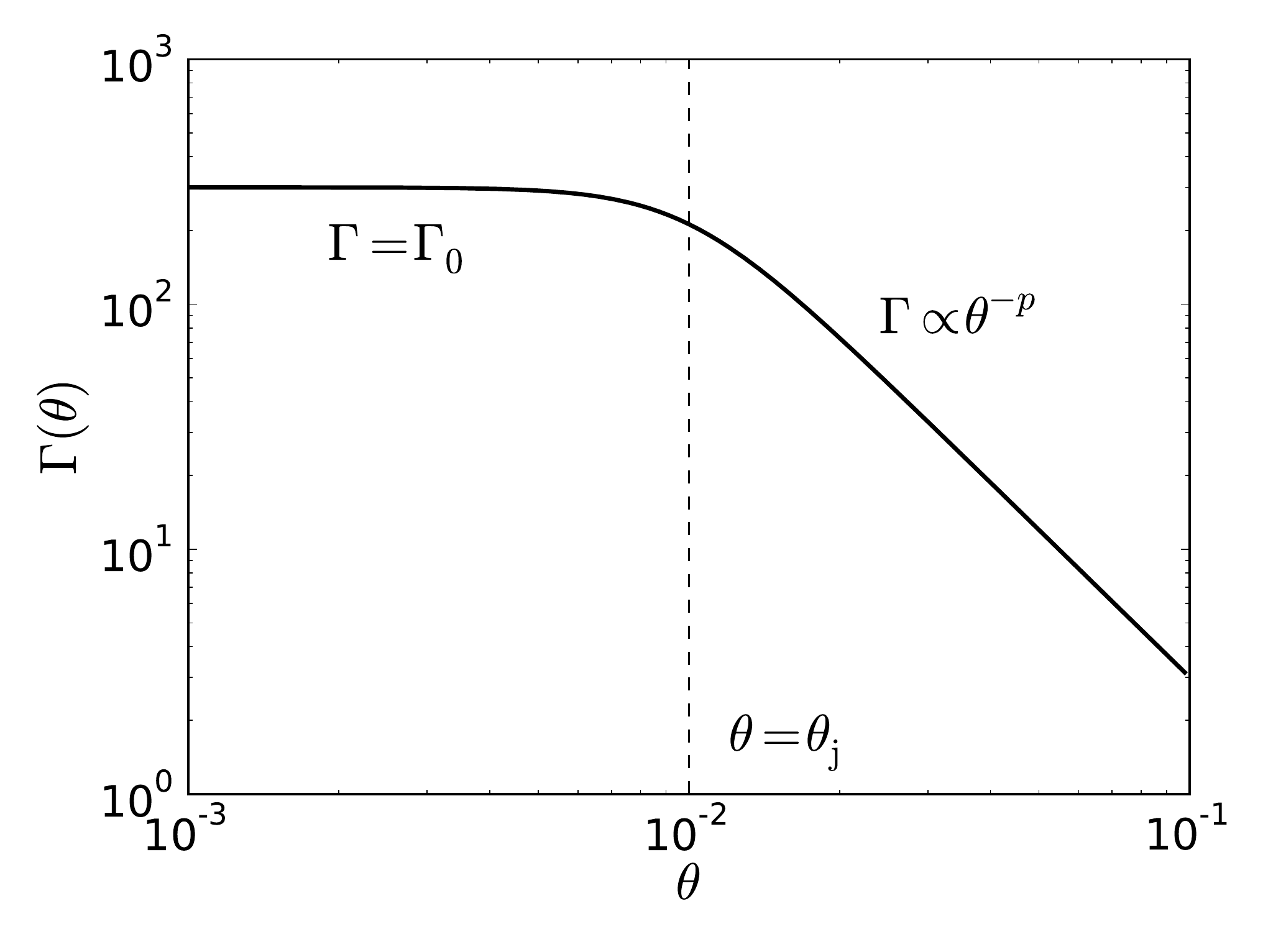}
\caption{An example Lorentz factor profile (equation \ref{eq:lf profile}). The Lorentz factor is approximately constant, $\Gamma \approx \gn$, in the jet core ($\theta < \tj$), while in the shear layer ($\theta > \tj$) the Lorentz factor scales approximately as a power law with angle, $\Gamma \propto \theta^{-p}$. In our model, $\gn$, $\tj$ and $p$ are free parameters. For this figure $\gn = 300$, $\tj \gn = 3$ and $p=2$ were used.}
\label{fig:lf profile}
\end{figure}

Motivated by the angular Lorentz factor profiles presented by \citet{ZhaWooMac:2003}, we assume that the angular profile of the baryon loading leads to a saturated Lorentz factor of the form

\begin{equation}
\Gamma(\theta) = \frac{\gn}{\sqrt{(\theta/\tj)^{2p} + 1}},
\label{eq:lf profile}
\end{equation}

\noindent where $\gn$, $\tj$ and $p$ are free model parameters. As the saturated Lorentz factor is inversely proportional to the baryon loading (equation \ref{eq:eta}), equation \ref{eq:lf profile} together with the assumed outflow luminosity determines the baryon loading of the outflow. Equation \ref{eq:lf profile} implies that the Lorentz factor is approximately constant, equal to $\gn$, in the jet core ($\theta < \tj$) while the shear layer Lorentz factor scales approximately as a power law of the angle, $\Gamma \propto \theta^{-p}$ ($\theta > \tj$). A larger value of $p$ increases the steepness of the Lorentz factor gradient in the shear layer, which also decreases the angular width of the shear layer. The outer angle of the shear layer can be approximated as the angle where the Lorentz factor equals a few, $\ts \approx \tj (\gn/2)^{1/p}$ (where $\Gamma(\ts) = 2$ was used), and the width of the shear layer is $\ts - \tj \approx \tj[(\gn/2)^{1/p} - 1]$. The complete set of free model parameters is therefore $L, r_\mathrm{0}, \gn, \tj$ and $p$, as well as the observer viewing angle $\tv$, which is measured from the jet axis. An example Lorentz factor profile is shown in Figure \ref{fig:lf profile}.

An observer located at zero viewing angle sees deeper into the outflow than any other observer. For this observer, the photospheric radius is at a minimum along the LOS. By integrating equation \ref{eq:tau} from $r$ to infinity along the radial direction at $\theta=0$ and equating the resulting optical depth to unity, the radius of the photosphere along the LOS is found, $\rph(\tv=0) = L \sigma_\mathrm{T} / (8\pi m_\mathrm{p} c^3 \gn^3)$, where the Thomson scattering cross section, $\sigma_\mathrm{T}$, was used. The comoving temperature at this point in the outflow is $kT^\prime_\mathrm{ph} = 0.36 \, (\Gamma_\mathrm{0}/300)^{5/2} (L/10^{52} \, \mathrm{erg \, s^{-1}})^{-5/12} (r_\mathrm{0}/10^8 \, \mathrm{cm})^{1/6} \, \mathrm{keV}$ (while the observed temperature is Doppler boosted, $kT_\mathrm{ph}^\mathrm{ob} \approx 2\Gamma_\mathrm{0} kT_\mathrm{ph}^\prime$ for an on-axis observer). For non-zero viewing angles the photospheric radius is larger, and therefore the comoving temperature at the photosphere is lower. We therefore conclude that the electrons are cold (kinetic energies much less than $m_\mathrm{e}c^2$, where $m_\mathrm{e}$ is the electron mass) in all relevant regions of the jet, and the scattering is in the Thomson regime, justifying the use of $\sigma = \sigma_\mathrm{T}$.

\subsection{Polarization properties of the photospheric emission: qualitative discussion}
\label{subsect:compton}

Polarization is an inherent feature of the Compton scattering process. Thomson scattering of an initially unpolarized photon beam at an angle of $\pi/2$ results in a fully linearly polarized outgoing beam. The polarization vector of the outgoing beam is orthogonal to the plane defined by the incoming and outgoing photon directions. There is therefore an inherent potential for observing linearly polarized emission from environments dominated by scattering, such as the photosphere. No circular polarization is induced by scatterings in the Thomson regime, and even if the initial photon field carries some degree of circular polarization, this polarization component quickly dissapears within a few scatterings. Therefore, we expect to observe only linear polarization from the photosphere\footnote{This statement is equivalent to $\V=0$, where $\V$ is the Stokes parameter for circularly polarized emission.}.

A basic requirement for producing a polarized signal by Compton scattering is that the comoving photon distribution in the fluid element where the last scattering occurs is anisotropic. This is because an isotropic distribution scatter equally into all directions, and as there is no preferred direction in the scattered photon field, there is no preferred direction for emission to be polarized in. As a photon propagates freely along a straight line between successive scatterings in an expanding outflow, the lab frame angle between the photon momentum vector and the local velocity direction decreases. If the lab frame angle decreases, so does the local comoving frame angle. This provides a source of anisotropy to the photon field. On the other hand, scattering reduces the comoving anisotropy by re-randomizing photon propagation directions. Deep down in the outflow where the optical depth is large and the photon mean free path is much smaller than the distance to the center of the outflow, the comoving photon angle is changed very little between scatterings, and the local comoving photon field can be considered isotropic. However, close to the photosphere the mean free path is of the order of the photospheric radius, and the change in comoving angle between scatterings is significant, which results in beaming of the local comoving photon field in the direction of the local velocity field. Therefore, the local comoving photon field is anisotropic at the last scattering position, and the escaping emission is polarized. For a thorough discussion on the comoving intensity in a spherical outflow, see \citet{Bel:2011}.

While an anisotropic local comoving intensity is a necessary requirement for producing a polarized signal, one additional requirement for spatially unresolved sources is some way of breaking the rotational symmetry of the emitting region. Consider a simplified model of a spherical outflow with Lorentz factor $\Gamma$, where all photons propagate strictly radially (corresponding to maximum comoving anisotropy) before making their last scatterings and reaching the observer. Assuming the electrons are cold, the photons that scatter at a comoving angle of $\pi/2$ are fully linearly polarized. This scattering angle corresponds to an angle $1/\Gamma$ in the lab frame, and so the emission from a single fluid element located at angle $1/\Gamma$ from the LOS is fully polarized. The polarization vector of the emission is orthogonal to the plane defined by the local radial direction and the LOS. In a spherical outflow that is spatially unresolved, the distribution of simultaneously observed fluid elements is symmetric around the LOS. This causes the polarization present in the emission from individual elements to average out. Therefore, some asymmetry must be present in the emitting region in order for the observed emission to be polarized. Now let us consider a non-spherical outflow. For observers located off-axis, the shape of the photosphere, and hence the shape of the emitting volume of the outflow, is not symmetric around the LOS. This provides a source of asymmetry, from which observed polarization emerges.

In order to qualitatively describe the observed polarization signal in photospheric emission, let us make a separation between the photons emitted by the jet core ($\theta < \tj$) and the shear layer ($\tj < \theta < \ts$). We start by describing the consequence of observing only the jet core. For observers located at $\tv \lesssim \tj - 1/\gn$, the emission appears to originate in a spherical outflow of Lorentz factor $\gn$, and no polarization is expected.

For larger viewing angles the observed photospheric radius of the core increases with viewing angle, because the angle between the propagation direction of the photons emitted towards the observer and the local velocity field increases. This has several consequences for the observed polarization. First, the observed flux decreases, since the jet core emission is beamed along the outflow propagation direction. Second, the peak energy of the observed emission decreases due to the lower Doppler boost of the observed photons. Third, the anisotropy of the comoving photon field at the last scattering position is increased. Fourth, the observed emitting region asymmetry around the LOS increases. The latter two points increase the polarization signal, while the former two points indicate possible correlations between the observed flux, average photon energy and polarization degree. Therefore, even for a top-hat jet we expect polarized emission at viewing angles $\tj - 1/\gn \lesssim \tv \lesssim \tj + 1/\gn$.

Including emission from the shear layer modifies the observed spectrum to a non-thermal shape, while decreasing the asymmetry of the emitting region. When the observer is located at $\tj \lesssim \tv \lesssim \ts$, the last scattering positions of the shear layer photons are more smoothly distributed around the LOS than those from the jet core, as the jet core points away from the observer. Decreasing the width of the shear layer increases the asymmetry of the observed emitting region. Therefore, a narrow shear layer in general leads to a greater observed polarized signal for a given viewing angle. The observed luminosity for observers located at viewing angles larger than $\ts$ can be neglected.

The typical observed photon energy of a shear layer photon is lower than the typical observed energy of a core photon for two reasons: the typical Lorentz factor is lower in the shear layer which results in a lower Doppler boost, and the photons lose more energy to adiabatic expansion before escaping the jet. Imposing a low energy cut on the observed emission (for instance, if the detector used to observe the outflow is only sensitive to photons above a given threshold) can therefore affect the asymmetry of the observed emitting region. In general, cutting away low energy photons is expected to somewhat increase the polarization degree, since photons from the shear layer are preferentially cut away.

The polarization vector of the emission integrated over the emitting region must point either orthogonal to, or lie in the plane defined by the jet axis and the LOS\footnote{Using the Stokes parameter definitions in \S \ref{sect:analytical} and forward, this statement is equivalent to $\U=0$.}. This is necessarily true for any jet with axial symmetry, as this leads to an observed emitting region with reflective symmetry above and below the plane, which in turn leads to only two orthogonal preferred directions of the emitting region. From here on, in order to simplify the discussion we choose to call the plane of symmetry the ``observer plane'', as all observers can be considered located within this plane.

\section{Simplified analytic treatment of the polarization properties of photospheric emission}
\label{sect:analytical}

Below we demonstrate that polarization of several tens of degrees is a natural consequence of photospheric emission from structured jets. In order to do this, we consider a toy model which takes into account emission from both the jet core and the shear layer. While the full treatment of radiative transfer is considered in \S \ref{sect:MC}, as discussed below the simplified model is a good approximation when viewing narrow jets ($\tj \gn \lesssim few$) at small viewing angles ($\tv \gn \lesssim few$). We note that the results of this section are obtained without the assumption of any particular outflow dynamics, and is therefore generally applicable to different astronomical objects.

Consider the scenario of a stationary, axisymmetric jet, pointing at an angle $\tv$ from the LOS of the observer. The number of photons streaming past the radius $\rph = \rph(\theta)$ per unit time, within the solid angle $\mathrm{d}\Omega$ as measured from the center of the outflow, is

\begin{equation}
\mathrm{d}\dot{N} (\theta, \rph) = \D{\dot{N}}{\Omega} (\theta, \rph) \mathrm{d}\Omega.
\end{equation}

\noindent Assuming isotropic scattering in the local comoving frame, the probability for a photon to make the last scattering into the comoving solid angle $\mathrm{d}\Omega^\prime_\mathrm{v}$ is $\mathrm{d}P = (1/4\pi) \mathrm{d}\Omega^\prime_\mathrm{v}$. Since the solid angle transforms as $\mathrm{d}\Omega_\mathrm{v} = D^{-2} \mathrm{d}\Omega^\prime_\mathrm{v}$ \citep{RybLig:1979}, the probability for a photon to scatter into the solid angle $\mathrm{d}\Omega_\mathrm{v}$ in the direction pointing towards the observer is

\begin{equation}
\mathrm{d}P (\theta, \tv) = \frac{D^2(\theta, \tv)}{4\pi} \mathrm{d}\Omega_\mathrm{v} (\tv),
\end{equation}

\noindent where $D = [\Gamma(1 - \beta\cos\tl)]^{-1}$ is the Doppler boost, $\tl$ is the angle between the local radial direction and the LOS, $\Gamma = \Gamma(\theta)$ is given by equation \ref{eq:lf profile} and $\beta \equiv \sqrt{1 - \Gamma^{-2}}$. By symmetry, the polar angle of a fluid element as measured from the LOS equals the angle between the LOS and the local radial direction. The photon rate emitted within the solid angle $\mathrm{d}\Omega_\mathrm{v}$ towards the observer from the solid angle $\mathrm{d}\Omega$ is then

\begin{equation}
\mathrm{d}^2\dot{N}^\mathrm{ob} (\theta, \tv, \rph) = \frac{D^2(\theta, \tv)}{4\pi} \D{\dot{N}}{\Omega} (\theta, \rph) \mathrm{d}\Omega \mathrm{d}\Omega_\mathrm{v} (\tv).
\end{equation}

The photon flux reaching the observer is partially polarized. We now aim to compute the observed polarization degree of photons emitted by all parts of the jet. This is accomplished by considering the polarization properties of the photons emitted by a local fluid element and integrating the contributions from all parts of the jet. We assume the solid angle extended by the emitting region from the center of the outflow to be confined within the angle of the outer part of the shear layer, $\ts$. We denote this solid angle by $\Omega_\mathrm{s}$. The total number of photons emitted per second and steradian from the jet towards the observer is then

\begin{equation}
\D{\dot{N}^\mathrm{ob}}{\Omega_\mathrm{v}} (\theta, \tv, \rph) = \frac{1}{4\pi} \int_{\Omega_\mathrm{s}} D^2 (\theta, \tv) \D{\dot{N}}{\Omega} (\theta, \rph) \mathrm{d}\Omega.
\label{eq:ph number}
\end{equation}

As discussed in \S \ref{subsect:compton}, the observed emission can only be polarized parallel or perpendicular to the observer plane. Therefore, the polarization properties of the photons are uniquely defined by the Stokes parameter ratio $\Q/\I$, where $\Q/\I = +1 \, (-1)$ indicates fully linearly polarized emission perpendicular (parallel) to the observer plane, the polarization degree is $\Pi = \lvert \Q \rvert /\I$ and $\I$ is the photon number intensity of the observed emission.

It was shown by \citet{Bel:2011} that emission propagating at a comoving angle of $\pi/2$ has $\Pi \approx 0.45$ close to, and above the photosphere in a spherical outflow. An unpolarized photon beam that Thomson scatters at an angle $\theta^\prime$ obtains the polarization degree $\Pi (\theta^\prime) = (1 - \cos^2\theta^\prime)/(1 + \cos^2\theta^\prime)$. We therefore approximate the polarization degree of emission making the last scattering into the comoving angle $\theta^\prime$ as $\Pi (\theta^\prime) \simeq 0.45 (1 - \cos^2\theta^\prime)/(1 + \cos^2\theta^\prime)$. Since the polarization degree of emission is invariant \citep{DeY:1966}, and the lab frame angle $\tl$ corresponds to a local comoving angle $\cos\theta^\prime = (\beta - \cos\tl)/(1 - \beta\cos\tl)$, the polarization degree of the observed emission from a local fluid element with the lab frame angle $\tl$ between the local radial direction and the LOS is

\begin{equation}
\Pi (\tl) \simeq 0.45 \frac{(1-\beta\cos\tl)^2 - (\cos\tl - \beta)^2}{(1-\beta\cos\tl)^2 + (\cos\tl - \beta)^2}.
\end{equation}

\noindent We define $\chi$ to be the angle between the observed projections on the sky of the local radial direction of a fluid element, and the jet axis\footnote{If a coordinate system is defined where the observer is located along the $z$-axis and the jet lies in the $x$-$z$ plane, $\chi = \pl$ where $\pl$ is the azimuthal angle as measured from the $x$-axis (see Appendix \ref{sect:appendix analytic solution}).}. Therefore, the contribution to the Stokes parameter ratio $\Q/\I$ from a local fluid element equals $\Pi \cos(2\chi)$, and we obtain

\begin{equation}
\frac{\Q}{\I} = \frac{\int_{\Omega_\mathrm{s}} D^2 (\mathrm{d}\dot{N}/\mathrm{d}\Omega) \Pi (\tl) \cos(2\chi) \mathrm{d}\Omega}{\int_{\Omega_\mathrm{s}} D^2 (\mathrm{d}\dot{N}/\mathrm{d}\Omega) \mathrm{d}\Omega}.
\label{eq:approx general}
\end{equation}

\noindent In our model, deep down in the outflow $\mathrm{d}\dot{N}/\mathrm{d}\Omega$ is constant with respect to angle from the jet axis (for $\theta \leq \ts$). A consequence of neglecting detailed radiative transfer is that $\mathrm{d}\dot{N}/\mathrm{d}\Omega$ stays constant as the emission approaches the photosphere. Therefore, the expression for the polarization degree simplifies to

\begin{equation}
\frac{\Q}{\I} = \frac{\int_{\Omega_\mathrm{s}} D^2 \Pi (\tl) \cos(2\chi) \mathrm{d}\Omega}{\int_{\Omega_\mathrm{s}} D^2 \mathrm{d}\Omega}.
\label{eq:approx int}
\end{equation}

\noindent Equations \ref{eq:ph number} and \ref{eq:approx int} may be solved numerically after the integration boundaries (corresponding to $\theta \leq \ts$) and the jet properties have been defined. This is done explicitly in Appendix \ref{sect:appendix analytic solution}.

The characteristic angle at which the optical depth changes for a photon that propagates within the jet core is $1/\gn$, the photon beaming angle. Therefore, the assumption of neglecting the treatment of opacity is not appropriate for viewing angles $\tv \gtrsim few/\gn$. The calculation presented above is therefore most applicable to the scenario of a jet with opening angle, shear layer width, and observer viewing angles all comparable to the photon beaming angle. These types of jets are also the ones from which the largest polarization degree is expected, since narrow jets with narrow shear layers maximize the asymmetry of the emitting region for a given viewing angle.

\begin{figure}
\includegraphics[width=\linewidth]{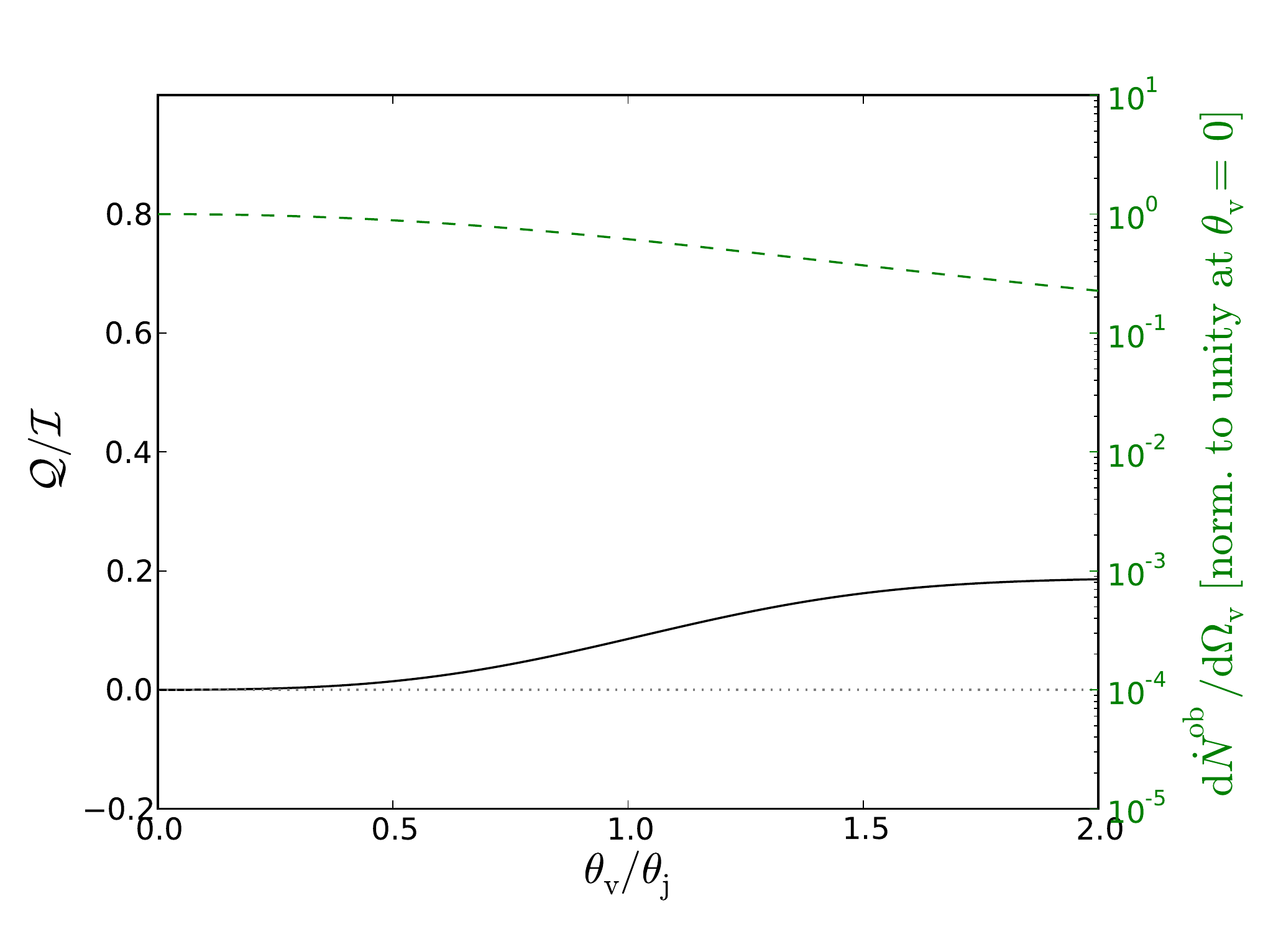}
\caption{The polarization properties (solid black line) and photon rate per solid angle (dashed green line, normalized to the photon rate at $\tv = 0$), of the observed photospheric emission as functions of the observer viewing angle obtained by solving equations \ref{eq:ph number} and \ref{eq:approx int} numerically. The parameters $\tj \gn = 1$ and $p=4$ were used. The gray dotted line indicates $\Q/\I = 0$ for reference.}
\label{fig:g100j001p4_poldeg_approx}
\end{figure}

Figure \ref{fig:g100j001p4_poldeg_approx} shows the numerical solutions to equations \ref{eq:ph number} and \ref{eq:approx int} for a jet with $\tj \gn = 1$ and $p=4$, within the viewing angle range $0 \leq \tv/\tj \leq 2$. As expected, the polarization degree for an on-axis observer equals zero. However, the polarization degree of emission at $\tv/\tj = 2$ reaches $20 \%$. The observed photon rate is also shown in Figure \ref{fig:g100j001p4_poldeg_approx}. It decreases slowly with viewing angle, indicating that the jet is observable up to angles of several times $\tj$. Computing the observed luminosity at a given viewing angle involves an estimation of the adiabatic energy losses for a local fluid element. This requires computing the surface of the observed photosphere, which requires integration of the optical depth between each fluid element and the observer. Due to the complexity of the problem, the observed luminosity was computed from a Monte Carlo simulation of the radiative transfer within the jet (see \S \ref{sect:MC}, \S \ref{sect:results}). At $\tv/\tj = 2$, the luminosity has dropped by close to an order of magnitude as compared to the luminosity for an on-axis observer. For the jet parameters considered here, the most likely observed viewing angle is $\gtrsim \tj$. Therefore, a large average polarization signal is expected if the polarization degree of the photospheric emission from a large sample of similar jets is measured.

\section{The radiative transfer code}
\label{sect:MC}

The toy model described in the previous section considers simplified radiative transfer. The Monte Carlo code described in this section was used to simulate the full radiative transfer in the jet. The transfer effects include bulk photon diffusion from the shear layer into the jet core and eventual polarization consequences of repeated scatterings. Furthermore, the complicated three-dimensional shape of the emitting region for off-axis observers is automatically taken into account.

The code tracks photons which undergo repeated scatterings in relativistically expanding plasmas. The propagation part of the code is designed to handle photon propagation in plasmas with angle dependent properties. Therefore, any non-thermal effects associated with photon propagation in shear layers are automatically considered. The scattering part of the code in the present version has been enhanced to include the treatment of photon polarization. Earlier versions of the code were used to study other aspects of photon propagation in regions of high optical depth and the resulting photospheric emission \citep{PeeWax:2004, PeeMesRee:2006b, Pee:2008, LunPeeRyd:2013}. In this section we describe the code in a qualitative way. A full description of the radiative transfer code appears in \citet{Pee:2008} and \citet{LunPeeRyd:2013}, while a description of our polarization treatment appears in Appendix \ref{sect:appendix simulation}.

The code makes use of dimensionless Stokes vectors, $\fvect{s} = (i, q, u, v)$, to represent the photon polarization properties. This formalism is convenient because of the additivity of Stokes parameters of incoherent ensembles of photons. By using the Stokes parameter formalism for single photons, which was originally defined using the intensities of incoherent photon beams, we allow for the polarization degree of each photon to vary between zero and unity. Therefore, after a scattering event, the outgoing photon carries the polarization properties that a beam of photons would have after scattering into the current direction, instead of being fully polarized at an angle which is drawn from the appropriate probability distribution. This treatment effectively removes a source of statistical uncertainty from the simulated scattering process. Since each photon in our simulation carries the same statistical weight, all Stokes parameters are normalized (divided by $i$) before being added together to form the Stokes parameters of the observed emission, $\fvect{S} = (\I, \Q, \U, \V)$. This method is similar to the methods used by \citet{BaiRam:1978} and \citet{JefKon:2011} in the context of solar flares, and \citet{Kra:2012} in the context of blazars.

There are three reference frames of importance to scattering problems: the lab frame, the local comoving frame and the electron rest frame. We define the lab frame as the reference frame in which the central engine of the outflow is stationary. The local comoving frame is the frame which is instantaneously comoving with the bulk outflow at a given location, which changes between scatterings. The electron rest frame is the frame which is stationary with respect to the specific electron on which the photon scatters, and is also different for each scattering event\footnote{Note that the electrons have a random Lorentz factor associated with the comoving temperature of the plasma, and therefore the electron rest frame differs from the local comoving frame.}. Between consequent scatterings a photon propagates along a straight line in the lab frame, which makes it the frame of choice for the propagation part of the code. The photon energy, direction and polarization properties after scattering is most easily obtained in the electron rest frame. Therefore, the code consists of an iterative process of propagating each photon a distance in the lab frame, followed by Lorentz transformations of the photon properties to the electron rest frame, via the local comoving frame. The scattering process is then performed, and the photon properties are transformed back to the lab frame to continue the propagation.

During a scattering process, the photon four-momentum and Stokes vector are transformed to the local comoving frame by consideration of the local velocity field at the scattering position. The electron distribution is assumed to be isotropic in the local comoving frame, with a Maxwellian energy distribution of the local comoving temperature given by equation \ref{eq:comoving temp}. The propagation direction and Lorentz factor of the scattering electron is drawn, after which the photon properties are transformed to the electron rest frame. The photon scattering direction is found, with a probability density distribution given by the polarization dependent Klein-Nishina cross section. After the scattered photon energy and polarization properties are computed, the photon four-momentum and Stokes vector is transformed back to the lab frame.

Between consecutive scattering events, the photon propagates freely along a straight line in the lab frame. In order to find the distance to the next scattering event, first the corresponding optical depth is drawn in the following way: the probability for a photon to scatter before propagating an optical depth $\tau$ is $P(\tau) = 1 - \exp(-\tau)$. Since $P(\tau)$ is a cumulative distribution, the corresponding probability density distribution from which we wish to draw the optical depth value is obtained by $f(\tau) = \mathrm{d}P(\tau)/\mathrm{d}\tau = \exp(-\tau)$. We define $u \equiv P(\tau)$ and solve for $\tau = \tau(u)$, which gives $\tau(u) = -\log(1-u)$. By drawing values of $u$ from a uniform distribution in the range $0 < u < 1$, values of $\tau$ are returned which conforms to the probability density distribution. The drawn optical depth is compared to the numerically integrated optical depth at a position infinitely far away in the photon propagation direction. If the drawn optical depth is larger, the photon is assumed to escape the outflow. Otherwise, the distance corresponding to the drawn optical depth is obtained. Since the outflow properties vary with angle to the jet axis, the optical depth between two points in space is obtained by numerical integration. A minimizing routine compares the numerically integrated optical depth with the drawn optical depth in an iterative process, where the end point of the numerical integration is modified until the acceptable tolerance is reached (the square of the optical depth difference is less than $10^{-6}$). After the corresponding distance is found, the photon location is updated to the new position and a scattering occurs. We consider the Thomson cross section in the optical depth calculation, because of the low photon energies involved.

In the present simulation, unpolarized photons ($\fvect{s} = (1, 0, 0, 0)$) are injected deep down in the outflow ($\tau = 20$ in the radial direction), where the comoving intensity can be considered isotropic. As the luminosity of the central engine is assumed to be isotropic, the initial photon position is chosen in an isotropic way. The comoving photon energy is drawn from a blackbody of the comoving outflow temperature at the injection point. The initial lab frame photon propagation direction is chosen such that the comoving intensity at the injection point is isotropic. The photon then propagates and scatters until it escapes the outflow. After the simulation process, the photons are binned in viewing angle, and the Stokes vectors are added to form the Stokes vector of the observed emission at any given viewing angle.


\section{Simulation results}
\label{sect:results}

\begin{figure}
\includegraphics[width=\linewidth]{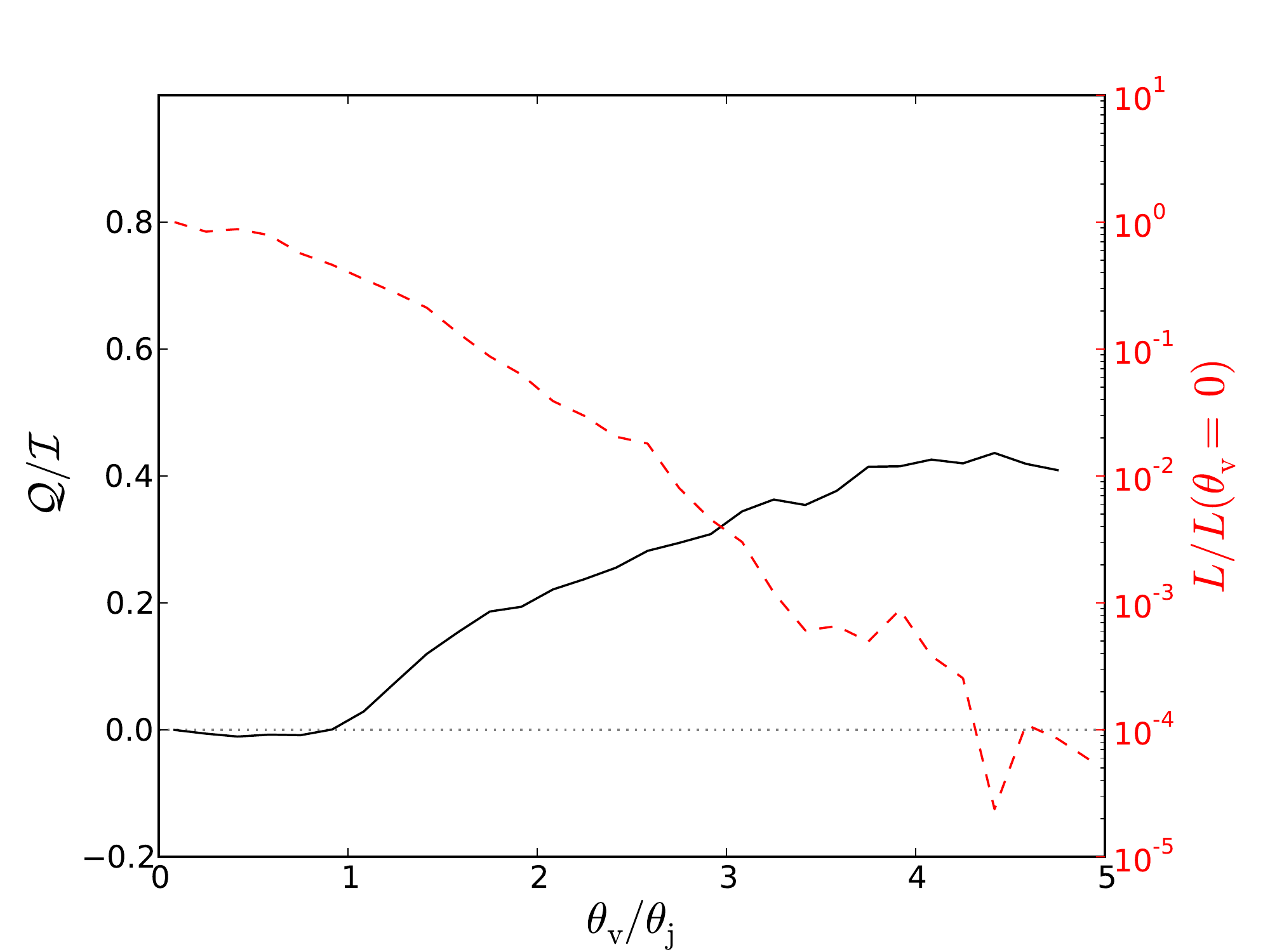}
\caption{The polarization properties (solid black line) and luminosity (dashed red line, normalized to the luminosity at $\tv = 0$) of the observed emission as functions of the observer viewing angle, for a narrow jet ($\tj \gn = 1$) with $p=4$. The gray dotted line indicates $\Q/\I = 0$ for reference. The polarization and luminosity are calculated using photons with $E / m_\mathrm{e} c^2 > 10^{-4}$. To avoid fluctuations due to low photon statistics, the polarization is only shown for viewing angle bins including more than $200$ simulated photons.}
\label{fig:g100j001p4_poldeg}
\end{figure}

\begin{figure}
\includegraphics[width=\linewidth]{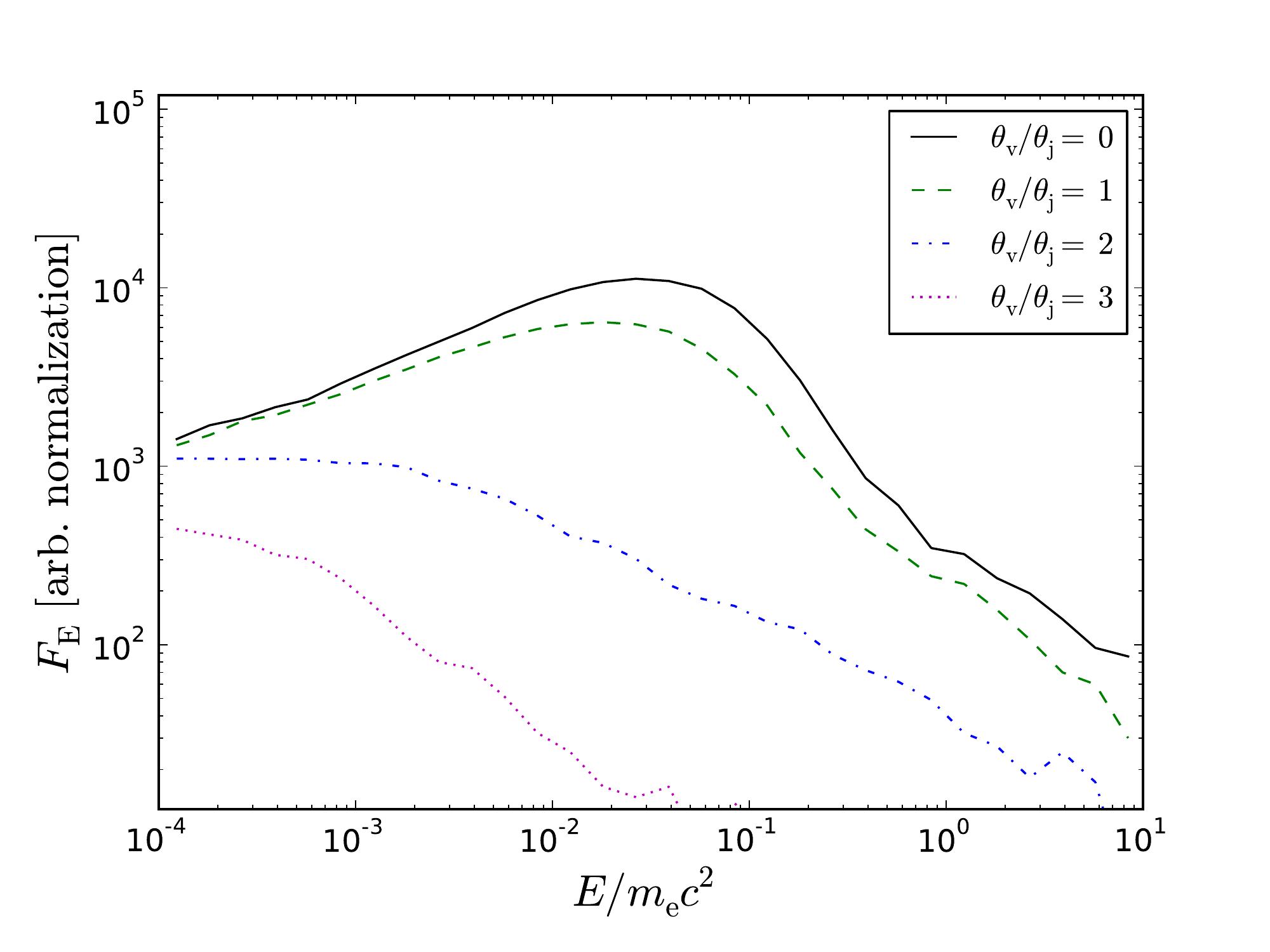}
\caption{Observed spectra from a narrow jet ($\tj \gn = 1$) with a narrow shear layer ($p=4$) observed at different viewing angles. The polarization degree of emission viewed at $\tv/\tj = 0$ and $\tv/\tj = 1$ is approximately zero, while the polarization degree at $\tv/\tj = 2$ is $\Pi \approx 20 \%$ and $\tv/\tj = 3$ is $\Pi \approx 40 \%$.}
\label{fig:g100j001p4_spec}
\end{figure}

\begin{figure}
\includegraphics[width=\linewidth]{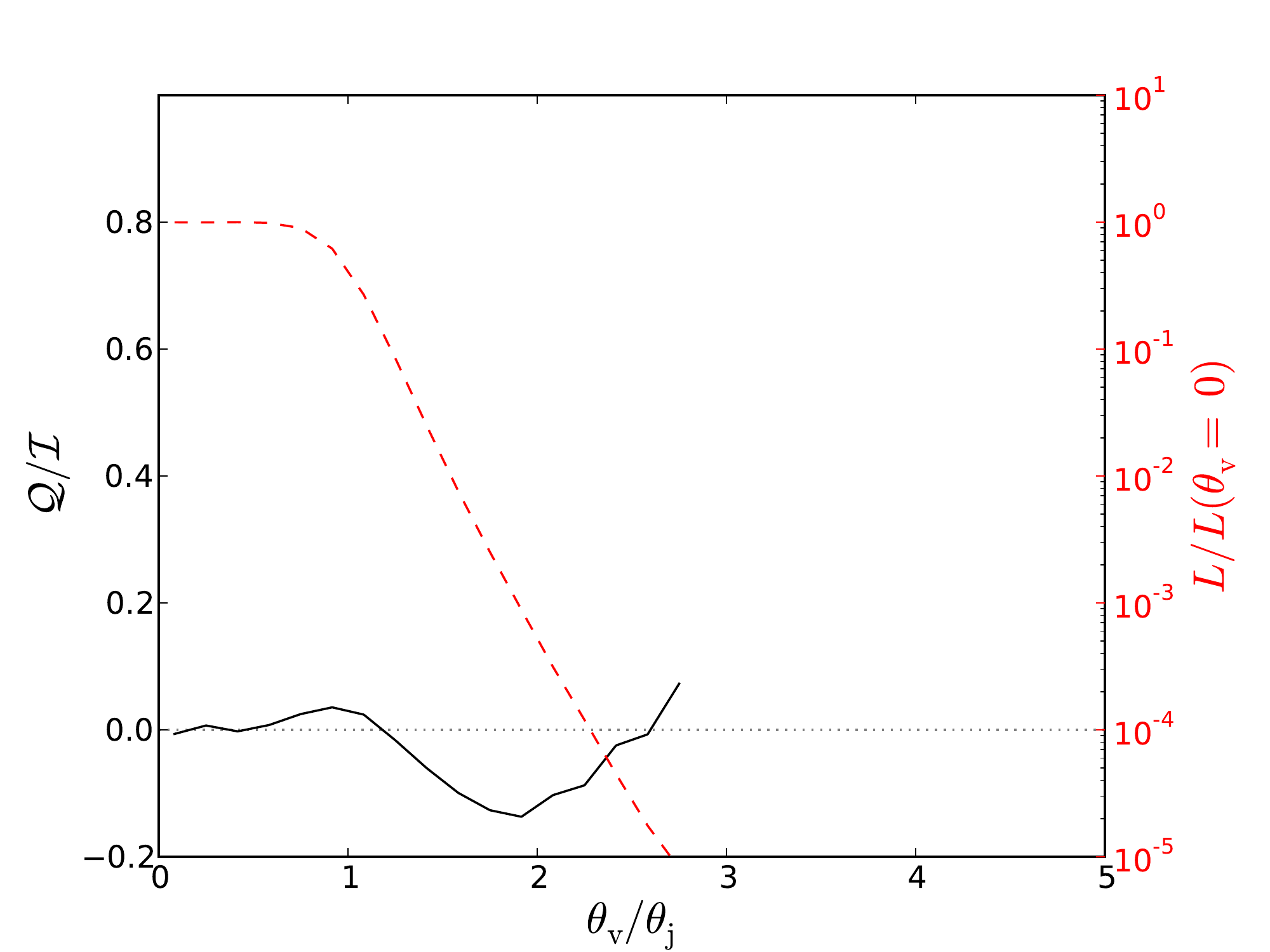}
\caption{The polarization properties (solid black line) and luminosity (dashed red line, normalized to the luminosity at $\tv = 0$) of the observed emission as functions of the observer viewing angle, for a wide jet ($\tj \gn = 10$) with $p=4$. The gray dotted line indicates $\Q/\I = 0$ for reference. The polarization and luminosity are calculated using photons with $E / m_\mathrm{e} c^2 > 10^{-4}$. The polarization is only shown for viewing angle bins including more than $200$ photons.}
\label{fig:g100j010p4_poldeg}
\end{figure}

\begin{figure}
\includegraphics[width=\linewidth]{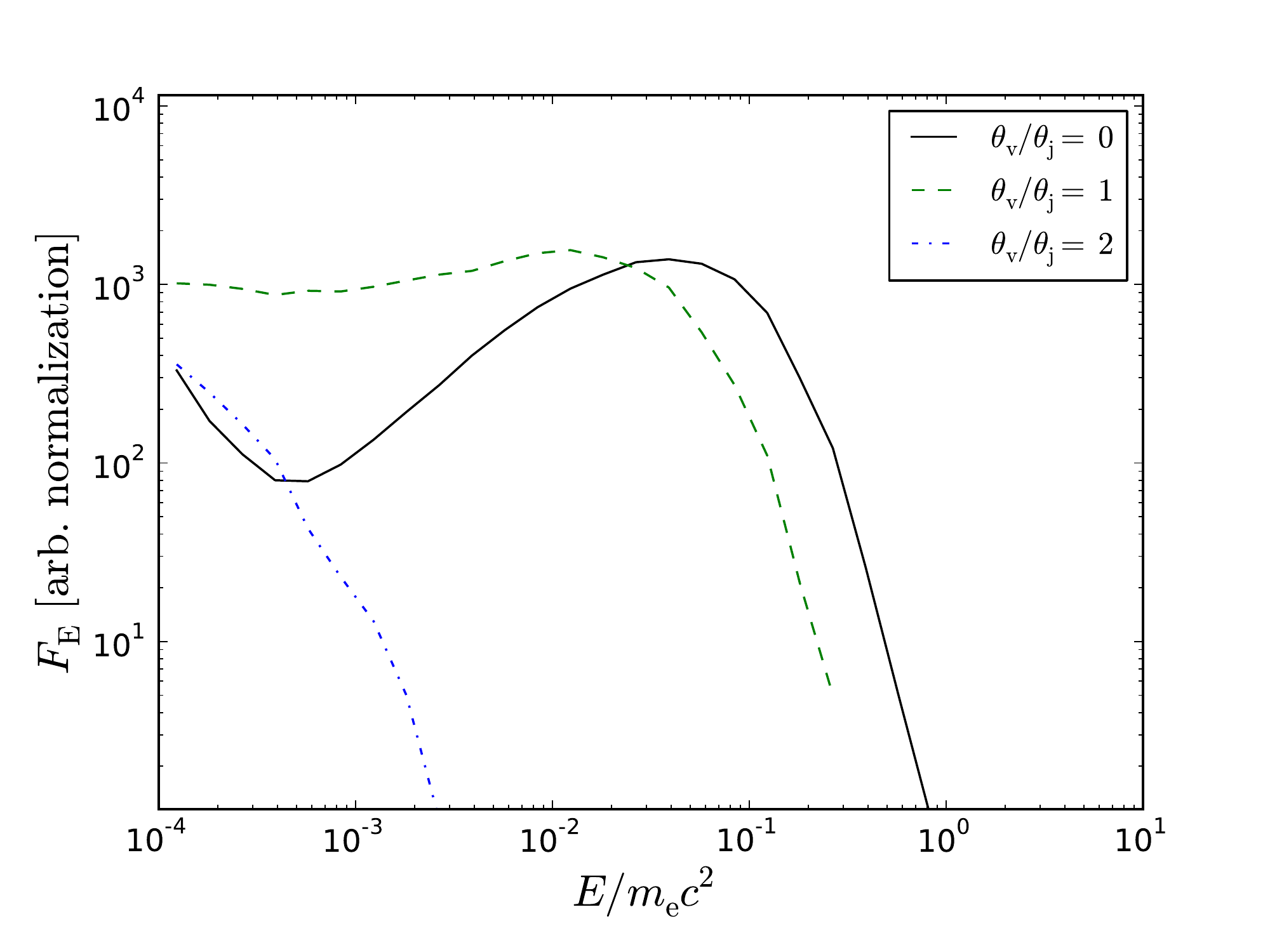}
\caption{Observed spectra from a wide jet ($\tj \gn = 10$) with $p=4$ observed at different viewing angles. The polarization degree of emission viewed at $\tv/\tj = 0$ is zero, while the polarization degree of emission viewied at $\tv/\tj = 1$ is $\Pi \approx 3 \%$ and $\tv/\tj = 2$ is $\Pi \approx 10 \%$.}
\label{fig:g100j010p4_spec}
\end{figure}

\begin{figure}
\includegraphics[width=\linewidth]{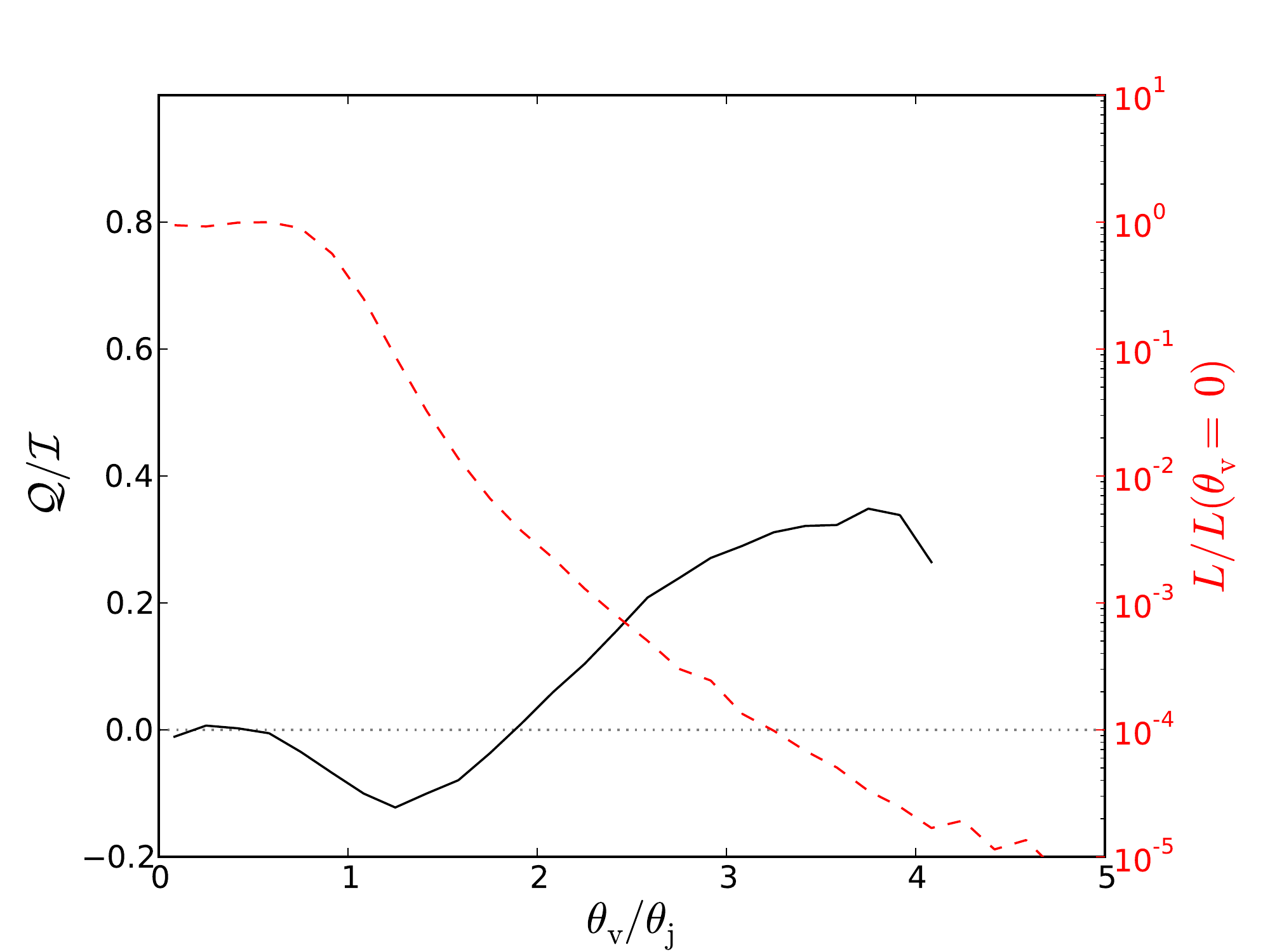}
\caption{The polarization properties (solid black line) and luminosity (dashed red line, normalized to the luminosity at $\tv = 0$) of the observed emission as functions of the observer viewing angle, for a jet of intermediate width ($\tj \gn = 3$) with $p=4$. The gray dotted line indicates $\Q/\I = 0$ for reference. The polarization and luminosity are calculated using photons with $E / m_\mathrm{e} c^2 > 10^{-4}$. The polarization is only shown for viewing angle bins including more than $200$ photons.}
\label{fig:g100j003p4_poldeg}
\end{figure}

In Figures \ref{fig:g100j001p4_poldeg} - \ref{fig:g100j003p4_poldeg} we present the results obtained from simulating the radiative transfer in the structured jets described in \S \ref{sect:model}. Typical central engine parameters characterizing GRBs were used: $L = 10^{52} \, \mathrm{erg \, s^{-1}}$ and $r_\mathrm{0} = 10^8 \, \mathrm{cm}$. The same parameter space as explored in \citet{LunPeeRyd:2013} has been considered: all combinations of the parameters $\tj \gn = \{1, 3, 10\}$ and $p = \{1, 2, 4\}$. As shown in \citet{LunPeeRyd:2013}, increasing $\gn$ increases the peak of the observed spectrum while keeping the spectral shape intact, as long as all other characteristic angles are decreased to keep the ratio with $1/\gn$ constant (i.e. all characteristic angles are rescaled). Numerically, it is more expensive to consider large Lorentz factors. While jets with different values of $\gn$ have been simulated, a value of $\gn = 100$ was used for producing the figures presented here. In presenting the results, a typical width of a viewing angle bin was chosen to be $\sim \tj/10$.

A top-hat jet is only visible up to viewing angles $\tv \approx \tj + 1/\gn$. Assuming $\tj \gg 1/\gn$ and that all jets point in random directions with respect to the observer, the expectation value of the viewing angle is $\sim (2/3) \tj$. However, because of photons emitted from the shear layer, some of the jets considered here are still luminous at angles several times $\tj$. Depending on the jet properties, the most probable viewing angle can be significantly increased so that most jets are observed at $\tv \gtrsim \tj$. We therefore present the simulated results in the range $0 \leq \tv/\tj \leq 5$.

\begin{figure}
\includegraphics[width=\linewidth]{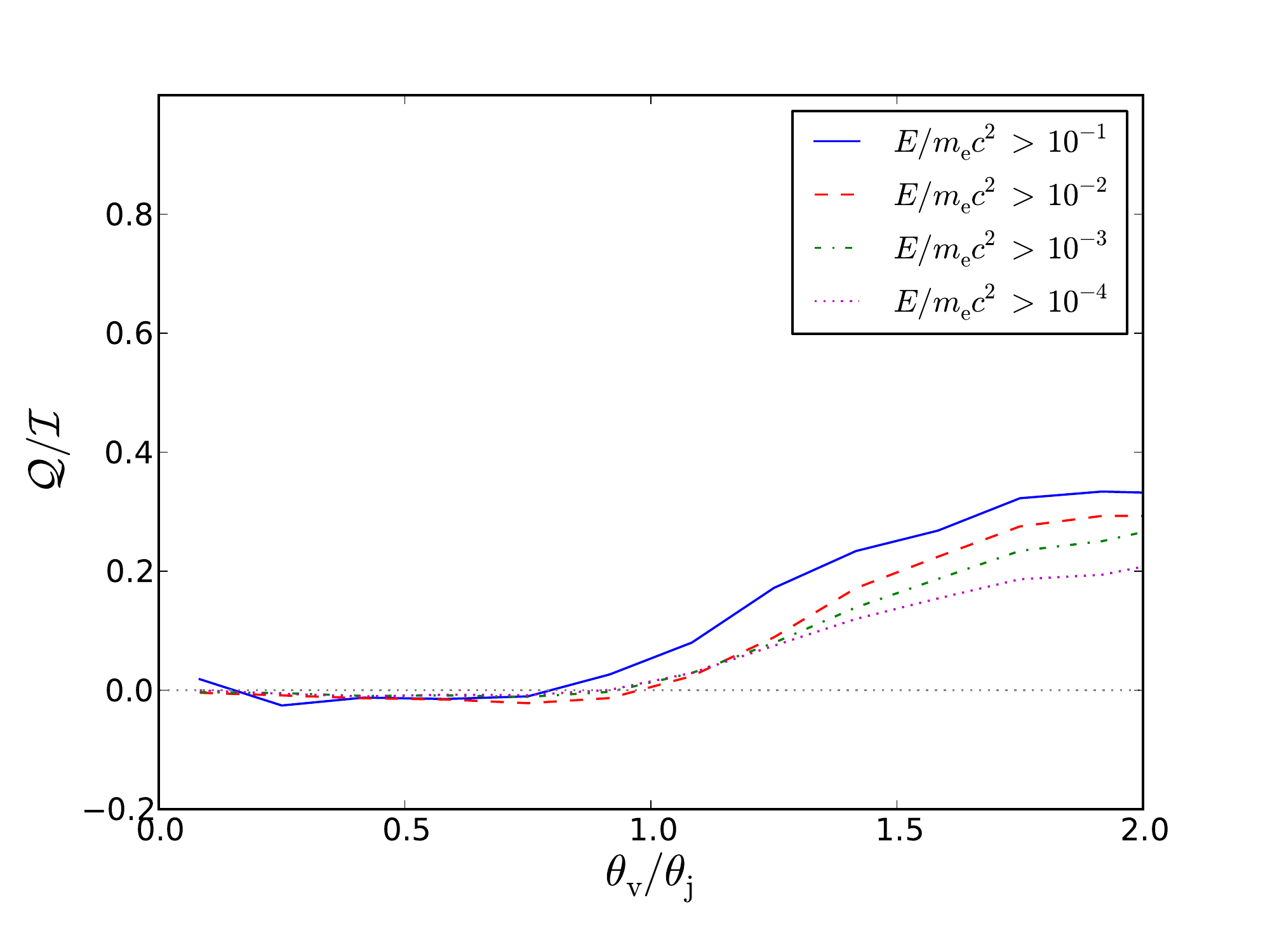}
\caption{The polarization properties of the observed emission with different low energy cuts, for a narrow jet ($\tj \gn = 1$) with $p=4$. The gray dotted line indicates $\Q/\I = 0$ for reference.}
\label{fig:g100j001p4_poldeg_eps}
\end{figure}

\begin{figure}
\includegraphics[width=\linewidth]{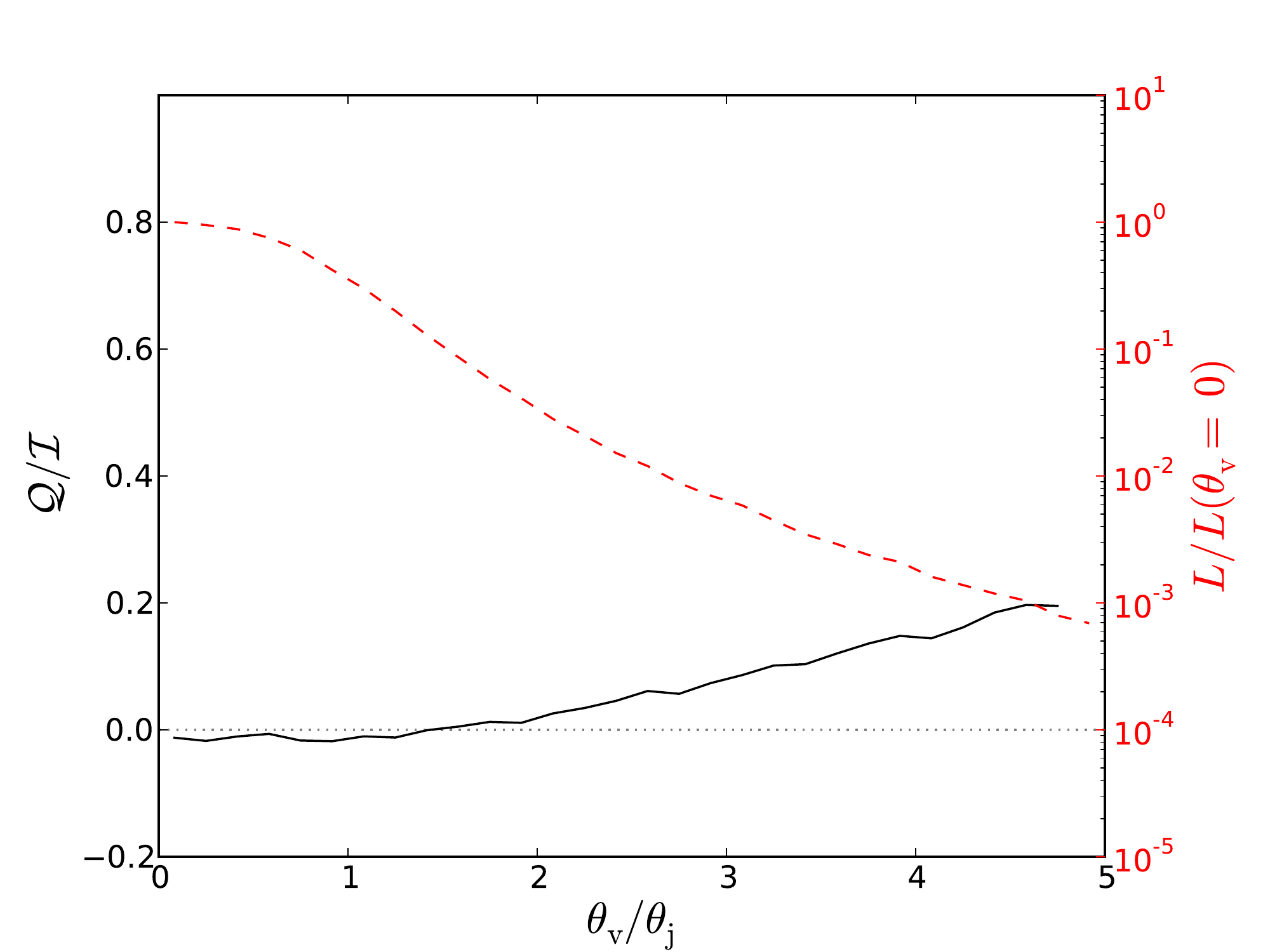}
\caption{The polarization properties (solid black line) and luminosity (dashed red line, normalized to the luminosity at $\tv = 0$) of the observed emission as functions of the observer viewing angle, for a narrow jet ($\tj \gn = 1$) with $p=2$. The gray dotted line indicates $\Q/\I = 0$ for reference. The polarization and luminosity are calculated using photons with $E / m_\mathrm{e} c^2 > 10^{-4}$. The polarization is only shown for viewing angle bins including more than $200$ photons.}
\label{fig:g100j001p2_poldeg}
\end{figure}

The thermal peak of the spectrum of photons emitted from these types of jets may correspond to the peak energy observed in the prompt emission of GRBs. Usually, the observed spectrum extends a few order of magnitude above and below the peak energy. We chose to present the results in a similarly defined energy range. For the chosen outflow parameters, we keep photons with $E/m_\mathrm{e} c^2 > 10^{-4}$ where $E$ is the observed photon energy. After applying the energy and viewing angle cuts, the simulation that Figures \ref{fig:g100j001p4_poldeg}, \ref{fig:g100j001p4_spec} and \ref{fig:g100j001p4_poldeg_eps} are produced from had $6 \cdot 10^5$ photons remaining. For Figures \ref{fig:g100j010p4_poldeg} and \ref{fig:g100j010p4_spec} the corresponding number is $10^6$ photons, for Figure \ref{fig:g100j003p4_poldeg} it is $4.5 \cdot 10^5$ photons and for Figure \ref{fig:g100j001p2_poldeg} it is $5.2 \cdot 10^5$ photons.

We plot the Stokes parameter ratio $\Q/\I$ in the figures. This ratio fully characterizes the polarization signal, since $\U = \V = 0$ (see \S \ref{subsect:compton}). The polarization degree of the emission is $\Pi = \sqrt{\Q^2 + \U^2 + \V^2}/\I = \lvert \Q \rvert /\I$. A positive value of $\Q$ corresponds to emission polarized perpendicular to the observer plane, while a negative value of $\Q$ indicates emission polarized within the observer plane.

One of our major findings is that emission from the photosphere can be highly polarized. Figure \ref{fig:g100j001p4_poldeg} shows the viewing angle dependence of $\Q/\I$ and the observed luminosity of emission from a narrow jet ($\tj \gn = 1$) with a narrow shear layer ($p=4$). As seen in Figure \ref{fig:g100j001p4_poldeg}, the polarization degree reaches $\Pi \approx 20 \%$ at $\tv/\tj = 2$, $\Pi \approx 30 \%$ at $\tv/\tj = 3$ and $\Pi \approx 40 \%$ at $\tv/\tj \gtrsim 4$. This is because of the large asymmetry of the emitting region achieved by considering both a narrow jet and a narrow shear layer. The observed luminosity at $\tv/\tj = 3$ is approximately $2.5$ orders of magnitude less than at $\tv/\tj = 0$. This implies that for plausible GRB parameters, the outflow is expected to be observed at these angles. The polarization degree calculated using the approximate analytical expression in \S \ref{sect:analytical} fits very well with the numerical results for $\tv/\tj \lesssim 2$.

By assuming that all GRBs are produced by narrow jets with narrow shear layers, that the jets are observable above some minimum luminosity, and that the GRBs are pointing uniformly in random directions, we can estimate the probability to observe a GRB with a polarization degree larger than some minimum value by taking the ratio of the solid angle of the polarized emission to the total observable solid angle. Assuming GRBs are observable in three order of magnitudes in luminosity (which is similar to the range reported by \citealt{GhiGhiFir:2006}), we obtain $P(\Pi > 30 \%) \approx 0.15$, $P(\Pi > 20 \%) \approx 0.62$ and $P(\Pi > 10 \%) \approx 0.80$ from Figure \ref{fig:g100j001p4_poldeg}.

Photons from the shear layer of narrow jets significantly affect the observed spectrum of the emission. The photon index below the thermal peak is $-1 \lesssim \alpha \lesssim -0.5$ ($\mathrm{d} N / \mathrm{d} E \propto E^\alpha$) for $1 \lesssim p \lesssim 4$ (for a thorough discussion, see \citealt{LunPeeRyd:2013}). Furthermore, if the shear layer itself is narrow (i.e. comparable to $few/\gn$), a power law is expected above the thermal peak, resulting from repeated scatterings between regions with different Lorentz factor (see \citealt{LunPeeRyd:2013, ItoEtAl:2013} for details). The observed spectrum emitted from narrow jets with narrow shear layers therefore has a broken power law shape for all observable viewing angles. This is presented in Figure \ref{fig:g100j001p4_spec}.

The fact that a narrow shear layer results in both large polarization degrees and efficient Comptonization implies a potential correlation between the strength of the emission above the spectral peak and the polarization degree of the prompt emission. Both effects are largest for top-hat jets. The non-thermal, Componized photons are visible at $\tv \gtrsim \tj - 1/\gn$, while significant polarization arises at $\tv \gtrsim \tj$. This implies that if the observed emission is highly polarized, a tail of Comptonized photons should be observed above the thermal peak. However, this tail may be observed also at smaller viewing angles where the polarization degree is low (for a narrow jet, the tail is visible even for on-axis observers, for which the polarization averages out). The correlation could be used to test the hypothesis that the observed GRB emission above the spectral peak is due to Comptonization of photons in the shear layer.

We find that for wider jets, within the considered parameter space, the polarization degree is lower than for narrow jets. This is illustrated in Figure \ref{fig:g100j010p4_poldeg}, where $\tj \gn = 10$ and $p=4$ were used and the polarization degree peaks at $\Pi \approx 13 \%$. In our parameterization the width of the shear layer is proportional to $\tj$, and therefore a wider jet also have a wider shear layer as compared to $1/\gn$, which decreases the observed asymmetry of the emitting region. We expect a larger polarization degree from wide jets with narrower shear layers.

As shown in Figure \ref{fig:g100j010p4_poldeg}, the emission from wide jets is polarized either parallel, or perpendicular to the observer plane, depending on the viewing angle. This is a consequence of the shear layer not being visible to all observers. For a detailed discussion, see Appendix \ref{sect:appendix emitting region}.

An important finding is that for all viewing angles where a significant polarization degree is observed ($\Pi \gtrsim few \, \%$), the spectrum below the thermal peak has an index in the range $-1 \lesssim \alpha \lesssim -0.5$. This is illustrated for a jet with $\tj \gn = 1$, $p=4$ in Figures \ref{fig:g100j001p4_poldeg} and \ref{fig:g100j001p4_spec} and $\tj \gn = 10$, $p=4$ in Figures \ref{fig:g100j010p4_poldeg} and \ref{fig:g100j010p4_spec}. The underlying reason is that high degrees of polarization requires significant viewing angles, and for those viewing angles the shear layer is clearly observable.

Figure \ref{fig:g100j003p4_poldeg} shows the polarization properties of emission from a jet of intermediate width ($\tj \gn = 3$) with $p = 4$. The polarization degree peaks at $\Pi \approx 37 \%$ for large viewing angles, similar to narrow jets. At $\tv/\tj \approx 2.5$, where the luminosity is approximately three orders of magnitude below the luminosity at zero viewing angle, the polarization degree is $\Pi \approx 20 \%$.

Figure \ref{fig:g100j001p4_poldeg_eps} is similar to Figure \ref{fig:g100j001p4_poldeg} ($\tj \gn = 1$, $p = 4$), but different low-energy cuts has been imposed on the emission. As can be seen, cutting the lower energy photons slightly increases the polarization degree. This is expected, as photons from the shear layer are preferentially cut away, increasing the asymmetry of the observed emitting region (see discussion in \S \ref{subsect:compton}).

In general, lower values of $p$ correspond to wider shear layers and less asymmetry in the emitting region. The decrease in asymmetry causes the observed polarization to be lower than from similar jets with narrower shear layers, while also causing a slower decrease of the observed luminosity with viewing angle. Figure \ref{fig:g100j001p2_poldeg} shows the polarization of emission from a narrow jet ($\tj \gn = 1$) with $p = 2$. The luminosity has decreased by three orders of magnitude at $\tv/\tj \approx 5$, where the polarization degree is $\Pi \approx 20 \%$.

\section{Summary and discussion}
\label{sect:discussion}

In this work we have considered the polarization properties of photospheric emission from structured jets consisting of a highly relativistic core and a shear layer with angle dependent baryon loading and Lorentz factor. In this context, polarized emission is achieved as a viewing angle effect. The jet core Lorentz factor, $\gn$, the core opening angle, $\tj$, and the Lorentz factor gradient in the shear layer, $p$, are free model parameters. A simplified version of the problem is solved analytically without any assumptions on the outflow dynamics, while the full radiative transfer in optically thick, non-dissipative fireball jets is simulated using a Monte Carlo code. The scenario considered differs from previous works, that considered the polarization properties of emission originating from optically thin regions of a top-hat jet.

We show that, contrary to common expectations, the emission from the photosphere may be strongly polarized. In particular, emission from a narrow jet ($\tj \gn \approx 1$) with a narrow shear layer ($p=4$) has a polarization degree of $\Pi \approx 40 \%$ at viewing angles where the luminosity is approximately three orders of magnitude lower than for an observer at zero viewing angle. Assuming the jet is observable in three orders of magnitude in luminosity (similar to the actual range observed, \citealt{GhiGhiFir:2006}), the probability to observe a polarization degree larger than $30 \%, 20 \%$ or $10 \%$ from such a jet is $P(\Pi > 30 \%) \approx 0.15$, $P(\Pi > 20 \%) \approx 0.62$ and $P(\Pi > 10 \%) \approx 0.80$, and the spectrum appears highly non-thermal for all viewing angles.

Within the considered parameter space, and for all viewing angles where a significant polarization degree is observed ($\Pi \gtrsim few \, \%$), the spectrum below the thermal peak has an index in the range $-1 \lesssim \alpha \lesssim -0.5$ due to geometrical broadening \citep{LunPeeRyd:2013}. The model therefore predicts that GRBs with $\Pi \gtrsim few \, \%$ will have low energy photon indices within the given range.

Furthermore, jets with narrow shear layers produce a power law of photons above the thermal peak. Within the considered parameter space this effect is most pronounced for a narrow jet with a narrow shear layer ($\tj \gn = 1, p=4$). The observed spectrum then becomes a smoothly broken power law, similar to what is observed in many GRBs. As this type of jet provides a large asymmetry of the observed emitting region for off-axis observers, the emission is highly polarized. Therefore, the jets that produce broken power law spectra also produce highly polarized emission for most observers ($\approx$ tens of percent), while the jets that produce spectra more similar to the Planck spectrum produce emission with lower degrees of polarization for most observers ($\approx$ a few, up to about ten percent).

\subsection{General considerations of the polarization of photospheric emission}
\label{subsect:general cons}

As explained in \S \ref{subsect:compton}, two requirements have to be satisfied in order to produce polarized emission from a spatially unresolved outflow dominated by scattering: the comoving intensity at the last scattering positions must be anisotropic and the emitting region must be asymmetric around the LOS. In order for the comoving intensity to be anisotropic, the outflow must expand and not be in the radiation dominated regime \citep{Bel:2011}. In order for the emitting region to be asymmetric, the jet needs to have a lateral structure and be viewed off-axis. The polarization properties of the emission are not sensitive to fireball properties such as the isotropic equivalent luminosity or the size of the base of the jet, but they are sensitive to the lateral jet structure (i.e. $\tj$, $\gn$ and $p$).

In this work we consider photospheric emission from non-dissipative jets. If the electrons in the jet are heated, the peak energy of the observed spectrum is increased \citep{Gia:2012, Bel:2013}. However, heating has only a small effect on the transfer of photon number \citep{Bel:2011}. Therefore, the polarization results obtained here are expected to be valid for heated jets as well. The details of the shape of the observed spectrum may nevertheless be modified by heating (e.g. \citealt{PeeMesRee:2006}). In particular, if a significant amount of the dissipated energy goes into both accelerating electrons and generating magnetic fields, additional synchrotron emission may complicate the spectral shape and polarization properties of the emission.

The total luminosity per solid angle of the initial fireball may be different from what has been considered in this work. If the luminosity in the shear layer is lowered significantly the emitting region of the jet approaches a top-hat. Since including photons from the shear layer generally decrease the asymmetry of the emitting region (see \S \ref{subsect:compton} and Appendix \ref{sect:appendix emitting region}), slightly larger polarization degrees may be obtained in such cases. As photons emitted from the shear layer can significantly soften the observed spectrum through geometrical broadening, the exact details of the spectral shape below the peak energy may be affected by different assumptions on the angle dependence of the total fireball luminosity.

\subsection{Shifts of the polarization angle}
\label{subsect:polarization angle}

As discussed in \S \ref{subsect:compton}, the polarization vector of emission observed from a spatially unresolved, axisymmetric jet may only point in two different directions. One is given by the projection of the jet axis on the sky, while the other direction is perpendicular to the first. We find that the polarization angle measured by an observer located at a fixed viewing angle depends on the jet width ($\tj \gn$). Therefore, if the jet width changes with time, the observed polarization angle may change by $90^\circ$.

The polarization properties of the prompt emission of GRB 100826A was measured using the {\it GAP} instrument \citep{YonEtAl:2011}. The data was split into two time intervals of approximately similar length for separate analysis. The reported polarization degrees and polarization angles are $\Pi = 25 \% \pm 15 \%$ and $\phi = 159^\circ \pm 18^\circ$ for the first interval and $\Pi = 31 \% \pm 21 \%$ and $\phi = 75^\circ \pm 20^\circ$ for the second interval. The polarization angle shift of $\sim 90^\circ$, as well as the polarization degrees can both be explained in the context of photospheric emission from a variable jet. As an example, consider a jet with fixed $\tj$ but varying $\gn$. A transition from a narrow jet ($\tj \gn \approx 1$) to a wider jet ($\tj \gn \gtrsim 3$) can shift the polarization angle by $90^\circ$ for the majority of the observers (see Figures \ref{fig:g100j001p4_poldeg} and \ref{fig:g100j010p4_poldeg} for observers at $\tv/\tj \approx 1.5 - 2$).

\subsection{Comparison to synchrotron emission}

The spectral shapes allowed by photospheric emission and synchrotron emission are different. In general, synchrotron emission produce wide spectra that may be characterized as a smoothly broken power law within a limited energy band. There exist well known limits on the value of the photon index below the peak energy from basic synchrotron theory. If the electrons cool efficiently by emitting synchrotron radiation (fast cooling electrons) the low energy photon index $\alpha$ must be smaller than, or equal to $-3/2$. This is in conflict with the majority of observed GRBs (e.g. \citealt{GolEtAl:2012}). On the other hand, if the electrons lose most of their energy to adiabatic expansion (slow cooling), the limit is $-2/3$. This is still inconsistent with $\sim 1/3$ of observations, and the synchrotron efficiency problem gets even worse. Photospheric emission can reach values as hard as $\alpha = 1$ (the Rayleigh-Jeans index) for wide, radiation dominated jets observed on-axis. By simple considerations of the jet structure, the possible range of indices extends to $-1 \lesssim \alpha \leq 1$ \citep{LunPeeRyd:2013}. Additional non-thermal emission resulting from energy dissipation close to the photosphere, or integration of time-varying spectra may reduce $\alpha$ even further. In general, it is less challenging to broaden a narrow spectrum than to make an inherently wide spectrum narrower.

The polarization degree of synchrotron emission from GRB jets has been considered by several authors (e.g. \citealt{Gra:2003, GraKon:2003, NakPirWax:2003, LyuParBla:2003, Wax:2003, Tom:2013}). Large polarization degrees are possible when the magnetic field in the emitting region is ordered, perpendicular to the local expansion direction (the maximally anisotropic configuration) and the jet is wide. Consider the photon indices $\alpha$ and $\beta$, below and above the spectral peak respectively ($\mathrm{d} N / \mathrm{d} E \propto E^\alpha, E^\beta$). \citet{Gra:2003} reports polarization degrees of $\Pi \approx 30 \%$ for $\alpha = -1$ (typical low energy photon index in GRBs, inconsistent with fast cooling electrons) and $\Pi \approx 65 \%$ for $\beta = -2.5$ (typical high energy photon index). \citet{NakPirWax:2003} finds $\Pi \approx 45 \% - 50 \%$ for an ordered magnetic field, while polarization degrees up to $\Pi \approx 35 \%$ was obtained for a top-hat jet with a magnetic field that is random within the plane perpendicular to the local expansion direction. However, these synchrotron calculations were performed under maximally asymmetric conditions (perfectly ordered field, or top-hat jet). If the magnetic field is somewhat curved within the emitting region or if the jet has a lateral structure (also considered by \citealt{NakPirWax:2003}), the polarization degree of synchrotron emission decreases due to the lower asymmetry of the emitting region. As a comparison, we find polarization degrees in the range $0 \% \lesssim \Pi \lesssim 40 \%$ for photospheric emission from a structured jet.

\citet{Tom:2013} argues that synchrotron emission theories needs to invoke a patchy jet in order to explain the shift of $\sim \pi/2$ in polarization angle as observed in GRB 100826A. A shift of $\pi/2$ is easily explained by photospheric emission from a jet with variable $\gn$ (see \S \ref{subsect:polarization angle}). In fact, this shift is the only one that is allowed by photospheric emission from an axially symmetric jet, while a patchy jet could provide a shift of any angle. Furthermore, the brightness of GRB 100826A implies a high efficiency of the prompt emission, which is natural in photospheric models. In contrast, the presumably large efficiency presents an additional challenge for the synchrotron interpretation \citep{Tom:2013}.

Synchrotron models predict a direct correlation between the photon index and the polarization degree of the emission (e.g. \citealt{Gra:2003, Tom:2013}). Emission with a softer spectral index is more highly polarized than emission with a hard spectral index. According to synchrotron theory, the emission from GRBs with lower values of $\alpha$ have a larger polarization degree. Similarly, GRBs with lower values of $\beta$ also have a larger polarization degree. From purely geometrical considerations, a similar correlation for $\alpha$ (albeit with a different range of allowed values) and an opposite correlation for $\beta$ exists for photospheric emission. If a wide jet is observed on-axis, the spectrum is narrow (somewhat wider than the Planck spectrum). Increasing the viewing angle widens the low energy spectrum as the shear layer comes into view, and simultaneously increases the polarization degree of the emission. Therefore, in the context of the photospheric emission model considered here, GRBs with lower values of $\alpha$ are expected to be more highly polarized (similar to synchrotron predictions). If the shear layer is narrow, increasing the viewing angle will also lead to observation of a Comptonized power law of photons above the peak energy. Therefore, GRBs with larger values of $\beta$ are expected to be more polarized (opposite to synchrotron predictions). The correlation is less pronounced for narrow jets. This is because the shear layer is visible also for on-axis observers, which causes the spectrum to change less with increasing viewing angle.


\section*{Acknowledgments}
The Swedish National Space Board is acknowledged for financial support.


\bibliographystyle{mn2e}
\bibliography{refgrb}

\appendix

\section[]{Numerical integration of the analytical model}
\label{sect:appendix analytic solution}

\begin{figure}
\includegraphics[width=\linewidth]{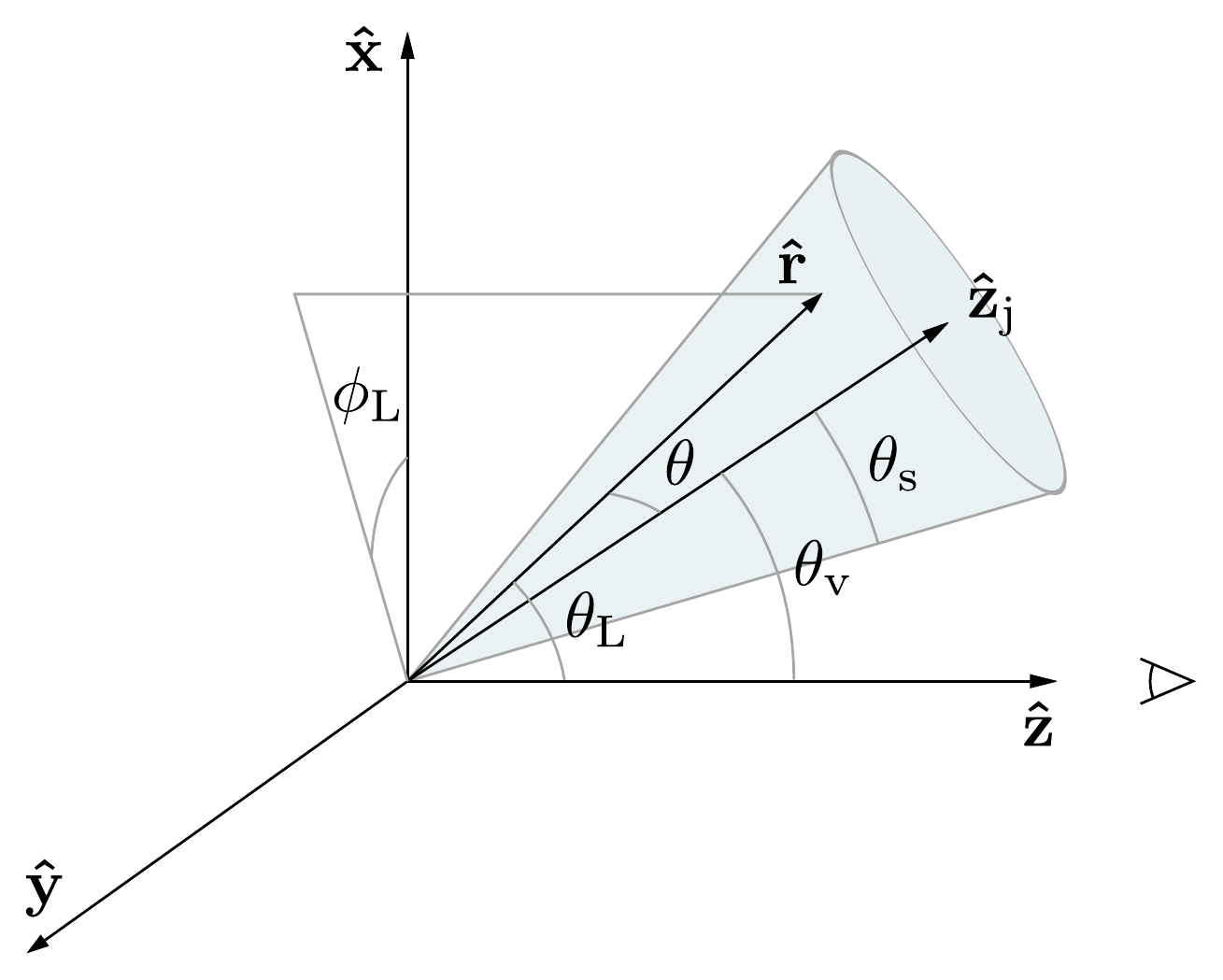}
\caption{The geometry considered in the simplified analytical calculation. The observer is located in the direction of the $z$-axis, while the jet points along the unit vector \vectsubhat{z}{j}. A general direction can be described by the unit vector \vecthat{r}, which has polar angle $\tl$ and azimuthal angle $\pl$.}
\label{fig:geometry}
\end{figure}

In order to perform the integrations in equations \ref{eq:ph number} and \ref{eq:approx int}, we define a right-handed coordinate system (shown in Figure \ref{fig:geometry}) in which the observer is located in the direction of the $z$-axis and the unit vector pointing along the jet axis is

\begin{equation}
\vectsubhat{z}{j} =
\begin{pmatrix}
\sin\tv\\
0\\
\cos\tv
\end{pmatrix}.
\end{equation}

\noindent Within this coordinate system, the $x-z$ plane defines the observer plane. A general position on the unit sphere, defined by the polar angle $\tl$ (measured from the $z$-axis) and the azimuthal angle $\pl$ (measured from the $x$-axis towards the $y$-axis), may then be expressed by the unit vector

\begin{equation}
\vecthat{r} =
\begin{pmatrix}
\sin\tl \cos\pl \\
\sin\tl \sin\pl \\
\cos\tl
\end{pmatrix}.
\end{equation}

\noindent By the definition of our coordinate system, the solid angle element can be expressed as $\mathrm{d}\Omega = \sin\tl \mathrm{d}\tl \mathrm{d}\pl$ while the angle $\chi$ between the projections on the sky of the local radial direction and the jet axis equals $\pl$. In order to make use of the Lorentz factor profile (equation \ref{eq:lf profile}) we need to express $\theta$ (the angle between $\vecthat{r}$ and the jet axis) as a function of $\tv$, $\tl$ and $\pl$: $\vectsubhat{z}{j} \cdot \vecthat{r} = \cos\theta = \sin\tl \cos\pl \sin\tv + \cos\tl \cos\tv$. The emitting region is confined to $\cos\theta \geq \cos\ts$, which gives a constraint on the variables of integration,

\begin{equation}
\cos\ts \leq \sin\tl \cos\pl \sin\tv + \cos\tl \cos\tv.
\label{eq:circ}
\end{equation}

\noindent The integration limits on $\tl$ and $\pl$ are found in the following way: the polar angle is limited by two constraints, $\pi \geq \tl \geq 0$ and $\tv + \ts \geq \tl \geq \tv - \ts$, which together define the integration limits on $\tl$. Since $\pl$ is measured from the $x$-axis, and the emitting regions on both sides of the observer plane possess reflective symmetry, the upper and lower limits on $\pl$ are of equal magnitude but opposite sign. Due to the reflective symmetry, the contributions to the observed emission from the jet on each side of the observer plane is equal, and we may consider the lower integration limit on $\pl$ to be zero, while multiplying the integrand by two.

If $0 \leq \tl \leq \ts - \tv$, the integration range for $\pl$ is $\pi \geq \pl \geq 0$, otherwise the upper limit on $\pl$ is determined by equation \ref{eq:circ} and the integration range equals

\begin{equation}
\arccos(\frac{\cos\ts - \cos\tl \cos\tv}{\sin\tl \sin\tv}) \geq \pl \geq 0,
\end{equation}

\noindent where $\sin\tl \sin\tv \neq 0$ has been assumed, which is true as long as $0 \leq \tl \leq \ts - \tv$ is false. The integrals in equations \ref{eq:ph number} and \ref{eq:approx int} may then be evaluated numerically.

\section[]{Detailed description of Lorentz transformations and scatterings of polarized photons}
\label{sect:appendix simulation}

Here we present details of the treatment of polarized radiative transfer used in our numerical code. A simulated scattering event consists of Lorentz transformations of the photon four-momentum and Stokes vector from the lab frame to the individual electron rest frame, followed by the actual scattering in the electron rest frame, and then transformations back to the lab frame. In this Appendix we describe how the code performs a Lorentz transformation of the photon properties as well as handling a scattering event.

The photon Stokes vector, $\fvect{s} = (i, q, u, v)^\mathrm{T}$, is always defined relative to a coordinate system in which the positive $z$-axis is parallel to the photon three-momentum\footnote{This implies that this vector can not be defined in an arbitrary coordinate system.}. We use the convention that $q/i = +1$ corresponds to full linear polarization parallel to the $y$-axis of the current coordinate system, $u/i = +1$ corresponds to full linear polarization in the direction pointing at $45^\circ$ from both the $x$-axis and the $y$-axis, and $v/i =+1$ corresponds to full left handed circular polarization.

All coordinate systems which the Stokes vector can be expressed in are related by a rotation around the photon three-momentum. Therefore, a matrix exists by which the Stokes vector can be expressed in a rotated coordinate system. The matrix which corresponds to a counter-clockwise rotation of the coordinate system an angle $\phi$ around the $z$-axis when the $z$-axis is pointing towards the reader equals \citep{McM:1961}

\begin{equation}
\vect{M}[\phi] = 
\begin{pmatrix}
1 & 0 & 0 & 0\\
0 & \cos2\phi & -\sin2\phi & 0 \\
0 & \sin2\phi & \cos2\phi & 0 \\
0 & 0 & 0 & 1
\end{pmatrix}.
\label{eq:M}
\end{equation}

\noindent The processes of Lorentz transformation and scattering of the Stokes vector simplify considerably when the Stokes vector is first expressed in appropriately rotated coordinate systems.

In our code, photons are allowed to propagate in any direction. While the photon four-momentum can be expressed in any coordinate system, such as the lab frame coordinate system $\coord$, in general the Stokes vector can not. Below we describe a general way to construct a coordinate system which the Stokes vector can be expressed in. We construct such a coordinate system by use of the photon three-momentum, $\vect{k}$, and any secondary vector which is not parallel to $\vect{k}$. The most convenient choice of the secondary vector depends on the actual situation considered. In order to keep the discussion general, we here denote the secondary vector by $\vect{A}$.

We define the $z$-axis of such a coordinate system to be parallel to $\vect{k}$, while the $y$-axis is defined to be parallel to $\vect{A} \times \vect{k}$. We denote this coordinate system by $\coordsup{Ak}$, and use the same superscript on vectors expressed in it. The $x$-axis of $\coordsup{Ak}$ is then obtained as the cross product of the $y$-axis and $z$-axis. In $\coordsup{Ak}$, both $\vect{A}$ and $\vect{k}$ lie in the $x-z$ plane.

When performing Lorentz transformations or scatterings of the Stokes vector it is useful to first express it in an appropriately rotated coordinate system. This is done by computing the angle of rotation, $\phi$, and use of the matrix $\vect{M}[\phi]$ in equation \ref{eq:M}. The angle of rotation equals the angle between the $y$-axes (or $x$-axes) of the current coordinate system of the Stokes vector, and the desired coordinate system (since they necessarily share the $z$-axis in order for the same Stokes vector to be expressable in both coordinate systems). The angle that needs to be supplied to $\vect{M}[\phi]$ in order to rotate the coordinate system from $\coordsup{Ak}$ to $\coordsup{Bk}$ is

\begin{equation}
\phi = -\mathrm{sign}[\vectsuphat{x}{Ak} \cdot \vectsuphat{x}{Bk}] \arccos(\vectsuphat{y}{Ak} \cdot \vectsuphat{y}{Bk}),
\label{eq:phi_rot}
\end{equation}

\noindent where $\mathrm{sign}[x]$ is a function that returns $+1$ if $x \geq 0$ and $-1$ otherwise, and $\coordsup{Bk}$ has been defined using the vectors $\vect{B}$ and $\vect{k}$. Changing coordinate systems for the Stokes vector from $\coordsup{Ak}$ to $\coordsup{Bk}$ is then performed by $\fvectsup{s}{Bk} = \vect{M}[\phi] \fvectsup{s}{Ak}$.

\subsection{General considerations of Lorentz transformation of the photon four-momentum and Stokes vector}

It was shown by \citet{DeY:1966} that the linear and circular polarization degrees of a statistical ensemble of incoherent photons are not affected by Lorentz boosts. Furthermore, \citet{Kra:2012} showed that there is a coordinate system from which the transformation of the Stokes vector is particularly simple: the coordinate system where the $y$-axis is perpendicular to both the photon three-momentum and the velocity vector of the frame into which the boost will be performed. If we choose to express the boosted Stokes vector in a similarly defined coordinate system in the boosted frame (i.e. the $y$-axis is perpendicular to both the boosted photon three-momentum and the direction of the velocity vector), then the Stokes parameter ratios $q/i$, $u/i$ and $v/i$ are left unchanged by the Lorentz transformation. While the intensity of a beam of photons is not invariant, we use the Stokes vectors for tracking the polarization properties of single photons, and therefore the vectors are kept normalized (i.e. divided by $i$) so that $i$ equals unity at all times. Thus the Stokes vector components, expressed in the respective coordinate systems of the two frames discussed above, are identical. The general Lorentz transformation from a given coordinate system in one frame, to another coordinate system in the boosted frame is then performed by two successive rotations of the Stokes vector by the matrix in equation \ref{eq:M}.

We now demonstrate in detail how to perform a Lorentz transformation of the photon four-momentum and the Stokes vector. For this example, we consider the transformation from the lab frame to the local comoving frame. The photon four-momentum is initially expressed in the lab frame coordinate system $\coord$, while the Stokes vector is expressed in $\coordsup{zk}$ (where the superscript $z$ refers to the unit vector pointing along the $z$-axis in $\coord$, although the choice of this particular vector is entirely arbitrary),

\begin{equation}
\fvect{P} =
\begin{pmatrix}
\epsilon \\
\vect{k}
\end{pmatrix},
\; \; \; \fvectsup{s}{zk} =
\begin{pmatrix}
i \\
q \\
u \\
v
\end{pmatrix}
\end{equation}

\noindent where $\epsilon$ is the photon energy in units of the electron rest mass. Our goal is to express the photon four-momentum in the comoving frame coordinate system $\coordsub{c}$ and the Stokes vector in $\coordsubsup{c}{zk_\mathrm{c}}$, where $\vectsub{k}{c}$ is the transformed photon three-momentum in $\coordsub{c}$ and the subscript $\mathrm{c}$ refers to the comoving frame.

The comoving frame moves with velocity $\boldsymbol\beta \equiv \vect{v}/c = (\beta_\mathrm{x} \; \beta_\mathrm{y} \; \beta_\mathrm{z})^\mathrm{T}$ with respect to the lab frame. Transformation of the four-momentum from $\coord$ to $\coordsub{c}$ is achieved by use of the matrix

\begin{equation}
\resizebox{\linewidth}{!}{$
\vect{\Lambda}[\boldsymbol\beta] = 
\begin{pmatrix}
\Gamma & -\Gamma \beta_\mathrm{x} & -\Gamma \beta_\mathrm{y} & -\Gamma \beta_\mathrm{z} \\
-\Gamma \beta_\mathrm{x} & 1 + (\Gamma-1) \beta_\mathrm{x}^2/\beta^2             & (\Gamma-1) \beta_\mathrm{x}\beta_\mathrm{y}/\beta^2     & (\Gamma-1) \beta_\mathrm{x}\beta_\mathrm{z}/\beta^2 \\
-\Gamma \beta_\mathrm{y} & (\Gamma-1) \beta_\mathrm{x}\beta_\mathrm{y}/\beta^2   & 1 + (\Gamma-1) \beta_\mathrm{y}^2/\beta^2               & (\Gamma-1) \beta_\mathrm{y}\beta_\mathrm{z}/\beta^2 \\
-\Gamma \beta_\mathrm{z} & (\Gamma-1) \beta_\mathrm{x}\beta_\mathrm{z}/\beta^2   & (\Gamma-1) \beta_\mathrm{y}\beta_\mathrm{z}/\beta^2     & 1 + (\Gamma-1) \beta_\mathrm{z}^2/\beta^2
\end{pmatrix},
$}
\label{eq:Lambda}
\end{equation}

\noindent where $\beta = \lvert \boldsymbol\beta \rvert$ and $\Gamma = [1-\beta^2]^{-1/2}$. The transformed four-momentum equals

\begin{equation}
\fvectsub{P}{c} = \vect{\Lambda}[\boldsymbol\beta]\fvect{P}.
\end{equation}

\noindent The transformed photon three-momentum, $\vectsub{k}{c}$, can then be obtained from $\fvectsub{P}{c}$.

Before transforming the Stokes vector, the angle $\phi$ which separates the $y$-axes of coordinate systems $\coordsup{zk}$ and $\coordsup{\beta k}$ is found by use of equation \ref{eq:phi_rot}, and the Stokes vector is rotated, $\fvectsup{s}{\beta k} = \vect{M}[\phi] \fvectsup{s}{zk}$. The transformation is then performed,

\begin{equation}
\fvectsubsup{s}{c}{\beta k_\mathrm{c}} = \fvectsup{s}{\beta k}.
\end{equation}

\noindent Finally, in order to express the Stokes vector in $\coordsubsup{c}{zk_\mathrm{c}}$ one additional rotation is needed. The angle $\tilde{\phi}$ needed to rotate from $\coordsubsup{c}{\beta k_\mathrm{c}}$ to $\coordsubsup{c}{zk_\mathrm{c}}$ is obtained by equation \ref{eq:phi_rot}, and the Stokes vector is rotated, $\fvectsubsup{s}{c}{zk_\mathrm{c}} = \vect{M}[\tilde{\phi}] \fvectsubsup{s}{c}{\beta k_\mathrm{c}}$.

\subsection{Scattering of the photon three-momentum and Stokes vector}

Here we describe how the code handles a scattering event. We assume that the photon vectors are already transformed to the electron rest frame before the calculation begins. As the scattering event involves rotations of the coordinate system in which the photon three-momentum is expressed, it is convenient to consider only the photon three-momentum instead of the four-momentum for the calculations. The coordinate system $\coord$ refers to the electron rest frame coordinate system, in which the incoming photon propagates in the direction specified by the angles $\tn$ and $\pn$. We use the subscript $\mathrm{0}$ to indicate photon properties before scattering, and no subscript after scattering. The goal of the calculation is then to find $\vect{k}$ and $\fvectsup{s}{zk}$ expressed in $\coord$ and $\coordsup{zk}$ respectively, from the initial $\vectsub{k}{0}$ and $\fvectsubsup{s}{0}{zk_0}$ expressed in $\coord$ and $\coordsup{zk_0}$.

The photon properties before scattering are

\begin{equation}
\vectsub{k}{0} = \epsilon_\mathrm{0}
\begin{pmatrix}
\sin\tn \cos\pn \\
\sin\tn \sin\pn \\
\cos\tn
\end{pmatrix},
\; \; \; \fvectsubsup{s}{0}{zk_0} =
\begin{pmatrix}
i_\mathrm{0} \\
q_\mathrm{0} \\
u_\mathrm{0} \\
v_\mathrm{0}
\end{pmatrix},
\end{equation}

\noindent We now wish to draw the scattering angles from the appropriate probability distribution. Since the photon propagates parallel to the $z$-axis of $\coordsup{zk_0}$ (by definition of the coordinate system), the photon three-momentum in this coordinate system is $\vectsubsup{k}{0}{zk_0} = (0 \; 0 \; \epsilon_\mathrm{0})^\mathrm{T}$. We therefore find the polar and azimuthal scattering angles in $\coordsup{zk_0}$, $\tsc$ and $\psc$, perform the scattering and rotate the scattered three-momentum back to $\coord$.

The two-dimensional probability density distribution from which the scattering angles are drawn is given by

\begin{equation}
\D{P}{\Omega} (\tsc, \psc) = \frac{1}{\sigma} \D{\sigma}{\Omega}(\tsc, \psc),
\end{equation}

\noindent where $\mathrm{d}\sigma/\mathrm{d}\Omega$ is the polarization dependent, differential Klein-Nishina cross section and $\sigma = \int_{4\pi} (\mathrm{d}\sigma/\mathrm{d}\Omega) \mathrm{d}\Omega$. The differential cross section is (e.g. \citealt{BaiRam:1978})

\begin{equation}
\begin{split}
& \D{\sigma}{\Omega} = \frac{r_\mathrm{0}^2}{2} \left(\frac{\epsilon}{\epsilon_\mathrm{0}}\right)^2 \times \\ & \left\{\frac{\epsilon_\mathrm{0}}{\epsilon} + \frac{\epsilon}{\epsilon_\mathrm{0}} - \sin^2 \tsc \left(1 - (q/i)\cos 2\psc + (u/i)\sin 2\psc\right)\right\},
\end{split}
\end{equation}

\noindent where $r_\mathrm{0}$ is the classical electron radius, $\epsilon = \epsilon_\mathrm{0}/[1 + \epsilon_\mathrm{0}(1 - \cos\tsc)]$ is the photon energy after scattering and averaging over isotropic electron spin has been assumed. After $\tsc$ and $\psc$ have been drawn, the outgoing photon three-momentum in $\coordsup{zk_0}$ is

\begin{equation}
\vectsup{k}{zk_0} = \epsilon
\begin{pmatrix}
\sin\tsc \cos\psc \\
\sin\tsc \sin\psc \\
\cos\tsc
\end{pmatrix}.
\end{equation}

\noindent We express the three-momentum in $\coord$ by use of a rotation matrix of the general form,

\begin{equation}
\vect{R}[\theta, \phi] = 
\begin{pmatrix}
 \cos\theta\cos\phi & -\sin\phi & \sin\theta\cos\phi \\
 \cos\theta\sin\phi &  \cos\phi & \sin\theta\sin\phi \\
-\sin\theta         &         0 &         \cos\theta
\end{pmatrix}.
\label{eq:R}
\end{equation}

\noindent The initial photon angles in $\coord$ are used as arguments to the rotation matrix, $\vect{k} = \vect{R}[\tn, \pn] \vectsup{k}{zk_0}$ ($\vect{R}[\theta, \phi]$ is constructed by multiplication of two rotation matrices that rotates the coordinate system around the $y$-axis and $z$-axis separately).

Scattering of the Stokes vector is performed by multiplication with the scattering matrix \citep{McM:1961},

\begin{equation}
\resizebox{\linewidth}{!}{$
\begin{split}
& \vect{T}[\theta, \epsilon_\mathrm{0}] = \frac{1}{2} r_{\rm 0}^2 \left(\frac{\epsilon}{\epsilon_\mathrm{0}}\right)^2 \times \\ &
\begin{pmatrix}
1+\cos^2\theta + (\epsilon_\mathrm{0} - \epsilon)(1-\cos\theta) & \sin^2\theta & 0 & 0 \\
\sin^2\theta & 1 + \cos^2\theta & 0 & 0 \\
0 & 0 & 2\cos\theta & 0 \\
0 & 0 & 0 & 2\cos\theta + (\epsilon_\mathrm{0} - \epsilon)(1-\cos\theta)\cos\theta
\end{pmatrix}
\end{split}
$}
\label{eq:T},
\end{equation}

\noindent where $\theta$ is the angle between the photon three-momentum vectors before and after scattering. The scattering matrix in the form presented above is applicable when the Stokes vector before scattering is expressed in $\coordsup{k k_0}$. The Stokes vector after scattering is expressed in $\coordsup{k_0 k}$. The angle required for rotating the Stokes vector coordinate system from $\coordsup{z k_0}$ to $\coordsup{k k_0}$ is $\psc$. The rotation is then performed, $\fvectsup{s}{k k_0} = \vect{M}[\psc] \fvectsup{s}{z k_0}$, and the Stokes vector is scattered,

\begin{equation}
\fvectsup{s}{k_0 k} = \vect{T}[\tsc, \epsilon_\mathrm{0}] \fvectsup{s}{k k_0}.
\end{equation}

\noindent The Stokes vector is then normalized. The angle $\phi$ required for rotating the Stokes vector coordinate system from $\coordsup{k_0 k}$ to $\coordsup{zk}$ is found by use of equation \ref{eq:phi_rot}, and the final rotation is performed, $\fvectsup{s}{z k} = \vect{M}[\phi] \fvectsup{s}{k_0 k}$.

\section[]{The asymmetry of the emitting region and the polarization angle}
\label{sect:appendix emitting region}

Close to the photosphere, the local comoving photon field is beamed along the radial direction (i.e. along the outflow propagation direction). This causes the scattered emission from a local fluid element to be polarized perpendicular to both the local radial direction and the outgoing photon three-momentum. The observed emission from a local fluid element is therefore polarized perpendicular to the projection of the local radial direction on the sky.

\begin{figure}
\includegraphics[width=\linewidth]{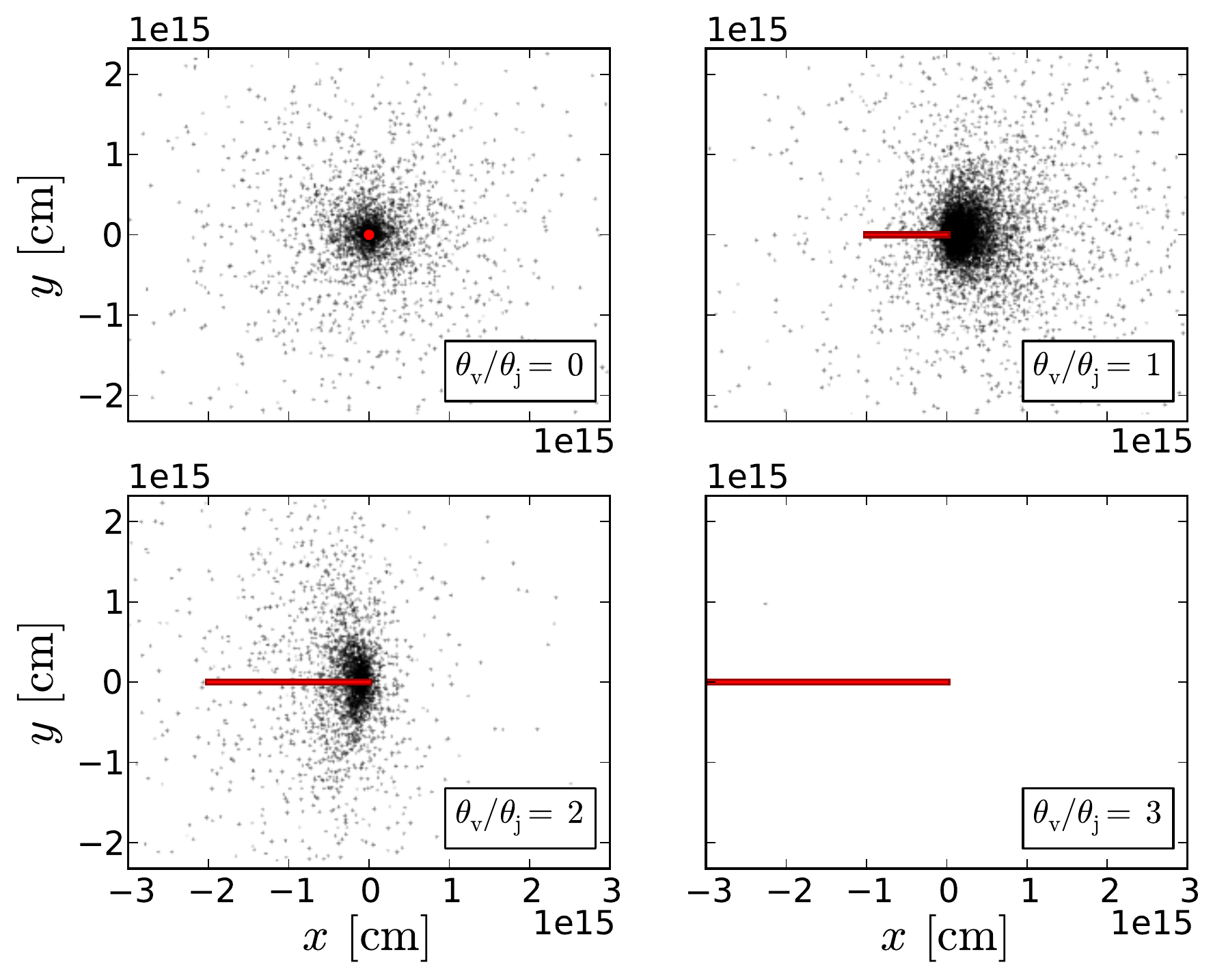}
\caption{Last scattering positions of simulated photons from a wide jet ($\tj \gn = 10$) with $p=4$, as projected onto the sky. The panels correspond to different viewing angles. The red line indicates the projection of the jet axis on the sky. The length of the projection is proportional to the viewing angle. The photons are from the simulation as presented in \S \ref{sect:results} (with parameters $\tj \gn = 10$ and $p=4$). At $\tv/\tj = 1$ photons from the shear layer starts entering the field of view (from the right side). This makes the distribution of the last scattering positions elongated away from the projection of the jet axis, which results in $\Q > 0$. At $\tv/\tj = 2$ the distribution has moved further towards the jet axis, causing it to be vertically elongated, resulting in $\Q < 0$.}
\label{fig:g100j010p4_scatter}
\end{figure}

\begin{figure}
\includegraphics[width=\linewidth]{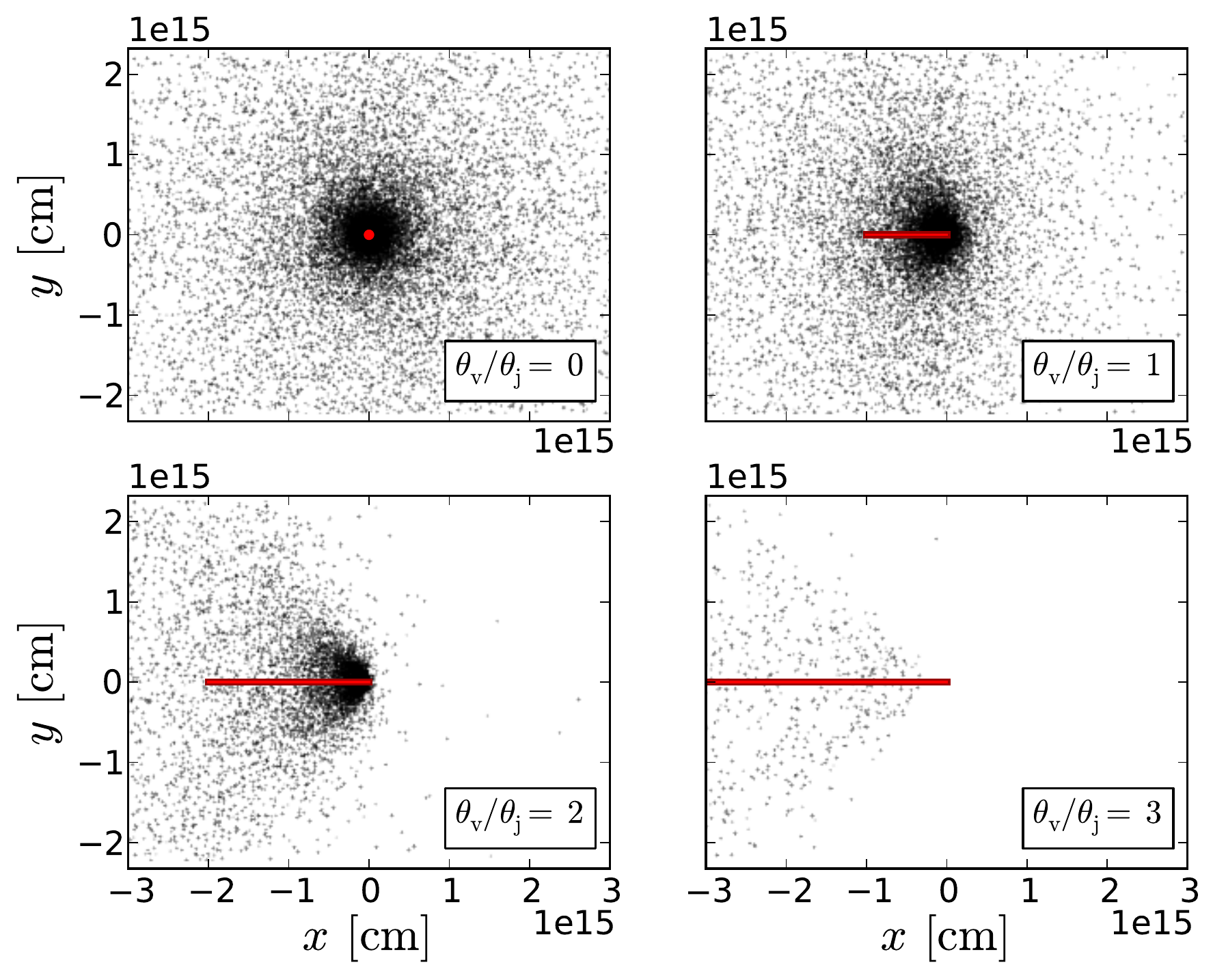}
\caption{Last scattering positions of simulated photons from a narrow jet ($\tj \gn = 1$) with $p=4$, as projected onto the sky. The panels correspond to different viewing angles. The red line indicates the projection of the jet axis on the sky. The length of the projection is proportional to the viewing angle. The photons are from the same simulation as presented in \S \ref{sect:results} (with parameters $\tj \gn = 1$ and $p=4$).}
\label{fig:g100j001p4_scatter}
\end{figure}

The emission from wide jets is polarized either parallel or perpendicular to the observer plane depending on the observer viewing angle (see Figures \ref{fig:g100j010p4_poldeg} and \ref{fig:g100j003p4_poldeg}). The explanation lies in the projected distribution of last scattering positions on the sky. Figures \ref{fig:g100j010p4_scatter} and \ref{fig:g100j001p4_scatter} show the projection of the last scattering positions of the observed photons onto the sky for a wide jet ($\tj \gn = 10$) and a narrow jet ($\tj \gn = 1$) respectively, as seen by observers located at different viewing angles. Let $\pl$ be the last scattering azimuthal angle (as previously defined in Figure \ref{fig:geometry}), measured from the projection of the jet axis on the sky. If the distribution of $\pl$ peaks close to $\pl = 0$ or $\pi$, fluid elements which contribute to $\Q > 0$ dominate and the emission is polarized orthogonal to the observer plane. If the distribution peaks at $\pl = \pi/2$ or $3\pi/2$, the emission is polarized within the observer plane ($\Q < 0$).

An observer located at zero viewing angle, observing a wide jet (Figure \ref{fig:g100j010p4_scatter}), do not see the shear layer. Only when the viewing angle becomes $\tv \approx \tj - 1/\gn$ does the shear layer come into view. At $\tv / \tj \approx 1$, the photons from the shear layer are more numerous than those from the jet core, as the emission from the jet core is beamed more strongly along the local radial direction. The distribution of $\pl$ then peaks away from the jet axis, at $\pl = \pi$, which leads to $\Q > 0$. At $\tv / \tj \approx 2$, the last scattering position of photons from the shear layer is centered around the LOS, but enlongated in the direction orthogonal to the observer plane. Consequently, the distribution of $\pl$ has two peaks at $\pl = \pi/2$ and $3\pi/2$, and the emission is polarized within the observer plane and $\Q < 0$. For even larger viewing angles the distribution of last scattering positions of all photons becomes elongated around $\pl = 0$, parallel to the projection of the jet axis on the sky. Therefore, the emission at these angles again obtain $\Q > 0$.

The situation is different for narrow jets (Figure \ref{fig:g100j001p4_scatter}), for which the shear layer is visible also for observers located at $\tv/\tj = 0$. Increasing the viewing angle leads to an increase in the projected anisotropy around the LOS and an increase in the observed polarization degree, while decreasing the observed flux. The distribution of $\pl$ peaks at $\pi$ for all observers, and the emission is therefore polarized perpendicular to the observer plane ($\Q > 0$) for all observers.

\label{lastpage}

\end{document}